# The SINS/zC-SINF survey of z~2 galaxy kinematics: evidence for gravitational quenching[1]

Genzel, R.[1,2,3], Förster Schreiber, N.M.[1], Lang, P.[1], S. Tacchella[4], Tacconi, L.J.[1], Wuyts, S.[1], Bandara, K.[1], Burkert, A.[5,1], Buschkamp, P.[1], Carollo, C.M.[4], Cresci, G.[6], Davies, R.[1], Eisenhauer, F.[1], Hicks, E.K.S.[7], Kurk, J.[1], Lilly, S.J.[4], Lutz, D.[1], Mancini, C.[8], Naab, T.[9], Newman, S.[3], Peng, Y.[4,13], Renzini, A.[8], Shapiro Griffin, K.[10], Sternberg, A.[11], Vergani, D.[12], Wisnioski, E.[1], Wuyts, E.[1] & Zamorani, G.[14]

[1] *Max-Planck-Institut für extraterrestrische Physik (MPE), Giessenbachstr.1, D-85748 Garching, Germany (genzel@mpe.mpg.de)*

[2] *Department of Physics, Le Conte Hall, University of California, Berkeley, CA 94720, USA*

[3] *Department of Astronomy, Campbell Hall, University of California, Berkeley, CA 94720, USA*

[4] *Institute for Astronomy, Department of Physics, Eidgenössische Technische Hochschule, ETH Zürich, CH-8093, Switzerland*

[5] *Universitäts-Sternwarte Ludwig-Maximilians-Universität (USM),* Scheinerstr. 1, München, D-81679, Germany

[6] *Istituto Nazionale di Astrofisica – Osservatorio Astronomico di Arcetri, Largo Enrico Fermi 5, I - 50125 Firenze, Italia*

[7] *Department of Physics and Astronomy, University of Alaska Anchorage, 3211 Providence Drive, Anchorage, AK, 99507, USA*

[8] *Osservatorio Astronomico di Padova, Vicolo dell'Osservatorio 5, Padova, I-35122, Italy*

[9] *Max-Planck Institute for Astrophysics, Karl Schwarzschildstrasse 1, D-85748 Garching, Germany*

---

[10] *Space Sciences Research Group, Northrop Grumman Aerospace Systems, Redondo Beach, CA 90278, USA*

[11] *School of Physics and Astronomy, Tel Aviv University, Tel Aviv 69978, Israel*

[12] *INAF Istituto di Astrofisica Spaziale e Fisica Cosmica di Bologna, Via P. Gobetti 101, 40129 Bologna, Italy*

[13] *Cavendish Laboratory, Room 908 Rutherford Building, JJ Thomson Avenue, Cambridge CB3 0HE, United Kingdom*

[14] *INAF Osservatorio Astronomico di Bologna, Via Ranzani 1, 40127 Bologna, Italy*

# Abstract

As part of the SINS/zC-SINF surveys of high-z galaxy kinematics, we derive the radial distributions of Hα surface brightness, stellar mass surface density, and dynamical mass at ~2 kpc resolution in 19 z~2 star-forming disks with deep SINFONI AO spectroscopy at the ESO VLT[1]. From these data we infer the radial distribution of the Toomre Q-parameter for these main-sequence star forming galaxies (SFGs), covering almost two decades of stellar mass ($10^{9.6}$ to $10^{11.5}$ $M_\odot$). In more than half of our SFGs, the Hα distributions cannot be fit by a centrally peaked distribution, such as an exponential, but are better described by a ring, or the combination of a ring and an exponential. At the same time the kinematic data indicate the presence of a mass distribution more centrally concentrated than a single exponential distribution for 5 of the 19 galaxies. The resulting Q-distributions are centrally peaked for all, and significantly exceed unity there for three quarters of the SFGs. The occurrence of Hα rings and of large nuclear Q-values is strongly correlated, and is more common for the more massive SFGs. While our sample is small and there remain substantial uncertainties and caveats, our observations are consistent with a scenario in which cloud fragmentation and global star formation are secularly suppressed in gas rich high-z disks from the inside out, as the central stellar mass density of the disks grows.

# 1. Introduction and Theoretical Background

Look-back studies have shown that most of the 'normal', massive star forming galaxies (SFGs) from z~0 to z~2.5 are located on or near a star formation 'main sequence' in the stellar mass ($M_*$) - star formation rate (SFR) plane, whose slope is near-universal (SFR ~ $M_*^{0.7...1}$) but whose amplitude, the specific star formation rate (sSFR=SFR/$M_*$), strongly changes with cosmic epoch (sSFR ~ $(1+z)^{2.9}$, Daddi et al. 2007, Noeske et al. 2007, Schiminovich et al. 2007, Rodighiero et al. 2010, 2011, Whitaker et al. 2012). As a result, the stellar buildup at early times is largely due to star formation on this main sequence.

The ionized gas kinematics of these SFGs (Genzel et al. 2006, 2008, Förster Schreiber et al. 2006, 2009, 2013, Law et al. 2009, Epinat et al. 2012, Vergani et al. 2012, Newman et al. 2013), as well as their rest-frame optical/UV brightness distributions (Wuyts et al. 2011b) suggest that 30-70% of the massive (log$M_*$ ($M_\odot$) >9.5) main-sequence star forming galaxies to z~2.5 are rotationally supported disks, albeit with large velocity dispersions, frequent perturbations due to minor mergers and highly clumpy and irregular appearances in UV/optical broad band imagery (Cowie et al. 1995, van den Bergh et al. 1996, Giavalisco, Steidel & Macchetto 1996, Elmegreen et al. 2004, 2009, Förster Schreiber et al. 2009, 2011a,b).

The first systematic studies of molecular gas in main-sequence SFGs from z~0 to z~3 find that the evolution of specific star formation rates above can be accounted for by corresponding changes in the molecular gas reservoirs, combined with a slowly changing depletion time scale of molecular gas to stars ($t_{depletion}$=$M_{mol\ gas}$/SFR~$t_0 \times (1+z)^{-\beta}$, with $t_0$~1.5±0.4 Gyrs and β~1±0.4, Tacconi et al. 2010, 2013, Genzel et al.

2010, Daddi et al. 2010a,b, Saintonge et al. 2011, 2012, 2013). High-z SFGs form stars rapidly, mainly because they are gas rich and globally unstable in their entire disks to gravitational fragmentation and star formation (Genzel et al. 2011).

These basic observational findings can be understood in a simple physical framework, in which global ('violent') gravitational instability and fragmentation in quasi-steadily fed, gas-rich disks create large, massive star forming clumps, which in turn drive turbulence through gravitational torques and stellar feedback (Noguchi 1999, Immeli et al. 2004 a,b, Bournaud, Elmegreen & Elmegreen 2007, Elmegreen, Bournaud & Elmegreen 2008, Genzel et al. 2008, Elmegreen 2009, Dekel et al. 2009a, Dekel, Sari & Ceverino 2009b, Bournaud 2010, Cacciato, Dekel & Genel 2012, Forbes et al. 2013). The most recent generation of cosmological galaxy evolution models and simulations find that the buildup of z>1 SFGs is dominated by smooth accretion of gas and/or minor mergers, and that stellar buildup at early times is largely due to in situ star formation (Kereš et al. 2005, 2009, Dekel & Birnboim 2006, Bower et al. 2006, Kitzbichler & White 2007, Ocvirk, Pichon & Teyssier 2008, Guo & White 2008, Dekel et al. 2009a, Davé, Finlator & Oppenheimer 2011, 2012). The large and quasi-steady gas accretion may plausibly build up early galaxy disks with a mass doubling time scale of ~0.5 Gyr at z~2 (Dekel et al. 2009a, Agertz, Teyssier & Moore 2009, Brooks et al. 2009, Ceverino, Dekel & Bournaud 2010). If the incoming material is gas rich, then global gravitational instabilities in these disks plausibly account for the large gas fractions and the star formation main sequence evolution inferred from the observations (Genel et al. 2008, Dekel et al. 2009b, Bouché et al. 2010, Davé et al. 2012, Lilly et al. 2013, Hirschmann et al. 2013).

Bulge formation in these early disks has traditionally been thought to occur in major mergers (e.g. Kauffmann & Haehnelt 2000, di Matteo, Springel & Hernquist 2005). The gravitational disk instability in early gas rich disks may open a second channel for bulge formation through internal radial gas transport. Star forming clumps and distributed gas in the disk are expected to migrate into the center via dynamical friction, viscosity and tidal torques, on a time scale of

$$t_{inspiral} \approx \left(v_c/\sigma_0\right)^2 t_{dyn}(R_{disk}) \sim 10\, t_{dyn}(R_{disk}) \sim t_{orb}(R_{disk}) < 0.5 \text{ Gyr} \quad (1),$$

where $t_{dyn}=R_{disk}/v_c$ and $t_{orb}=2\pi t_{dyn}$ are the mean disk dynamical and orbital time scales. The in-spiraling gas/stars may form a central bulge, and perhaps also a central massive black hole and a remnant thick disk (Noguchi 1999, Immeli et al. 2004 a,b, Förster Schreiber et al. 2006, Genzel et al. 2006, 2008, Elmegreen et al. 2008, Carollo et al. 2007, Dekel et al. 2009b, Bournaud, Elmegreen & Martig 2009, Ceverino et al. 2010). Inward radial transport depends strongly on $v_c/\sigma_0$. Importantly, since high-z disks are turbulent, the radial transport time scales are significantly smaller than the Hubble and gas depletion times, and are comparable to the orbital and mass-doubling time scales. In simulations the rate of mass inflow into the central region is comparable to the star formation rate in the disk (Dekel et al. 2013). The internal radial transport also redistributes angular momentum, resulting in higher angular momentum outer disks, relative to the inner stellar component, consistent with recent observations (Nelson et al. 2012).

A rotating, symmetric and thin gas disk is unstable to gravitational fragmentation if the Toomre Q-parameter (Toomre 1964) is below a critical value $Q_{crit}$. For a thin gas dominated disk in a background potential (of dark matter and an old stellar component) Q is related to the local gas velocity dispersion $\sigma_0$ (assuming isotropy), circular velocity $v_c$, epicyclic frequency $\kappa$ ($\kappa^2 = 2(v_c/R)^2 + (v_c/R) \, dv_c/dR$) and gas surface density $\Sigma_{gas}$ at radius $R$ via the relation (Wang & Silk 1994, Binney & Tremaine 2008, Escala & Larson 2008, Elmegreen 2009, Dekel et al. 2009b, Cacciato et al. 2012)

$$Q_{gas} = \frac{\kappa(R)\sigma_0(R)}{\pi G \Sigma_{gas}(R)} \qquad (2).$$

In the single component case $Q_{crit} \sim 1$. For a thick disk the surface gravity in the z-direction is lowered and the critical Q drops to $Q_{crit} \sim 0.67$ (Goldreich & Lynden-Bell 1965). The situation for multi-component thin or thick disks is more complicated, and depends on the Q-values of the individual components, as well as their velocity dispersions (Cacciato, Dekel & Genel 2012, Romeo & Falstad 2013). If the disk consists of molecular (H$_2$+He), atomic (HI+He) and stellar (*) components, $Q_{tot}^{-1} = Q_{H2}^{-1} + Q_{HI}^{-1} + Q_*^{-1}$ if all components have similar velocity dispersions, thus increasing the Q-thresholds for the individual components for the combined system to become critical. So for a thin disk of molecular gas and stars with the same $Q = Q_* = Q_{gas}$, the critical $Q_{gas}$ in the combined system becomes $Q_{crit,gas} \sim 2$. For a two component thick disk $Q_{crit,gas} \sim 1.32$.

Assuming that thick, high-z disks thermostat at marginal (in)stability, $Q \sim Q_{crit} \sim 0.67$-$1.3$, one finds from (2) with $\kappa = a \, v_c/R$

$$Q = \frac{av_c\sigma_0}{\pi RG\Sigma_{gas}} = a \times \frac{v_c^2 R/G}{\pi R^2 \Sigma_{gas}} \times \frac{\sigma_0}{v_c} = \frac{a}{f_{gas}} \times \frac{\sigma_0}{v_c} \qquad (3),$$

where *a* ranges between 1 (for a Keplerian rotation curve), 1.4 (for a flat rotation curve), and 2 (for a solid-body rotation curve) and $f_{gas}$ is the fraction of gas to the total mass in the disk (Genzel et al. 2008, 2011, Dekel et al. 2009b). For Q~1 $\sigma_0/v_c = f_{gas}/a$. This result and equation (1) show that the disk instability mechanism drives gas inward rapidly when the gas fraction is high, which is the case at z~1-3 but increasingly less so at lower redshifts.

If the radial gas transport discussed above builds up the central (mainly stellar) mass over a number of orbital time scales, and simultaneously the gas accretion rate into the disk slowly drops over cosmic time, or because the halo mass grows above $10^{11.6-12}$ M$_\odot$ (Rees & Ostriker 1977, Dekel & Birnboim 2006, Oczvirk et al. 2008, Dekel et al. 2009a), there should come a phase, depending on the efficacy of stellar feedback and radial gas transport, when Q in the central disk exceeds the critical value due to rotational shear (Hunter, Elmegreen & Baker 1998). The gravitational fragmentation process and the global disk instability may then shut off. This 'morphological' or 'gravitational' quenching mechanism (Martig et al. 2009, 2013) by itself cannot result in a permanent shutdown of star formation in the central disk, as long as gas is accumulating there due to radial transport. For this purpose either the radial transport into the center has to cease, or the accumulating but sterile gas needs to be removed, for instance by stellar or AGN feedback. Even if the gravitational quenching mechanism operates and the global Q exceeds the critical value, star formation may still occur in localized regions where dense, gravitationally bound clouds or cores form (see section 3.6); in essence gravitational quenching reduces the efficiency of star formation, and increases the molecular gas depletion time scale in the central parts of the disk. The global disk instability may also be rekindled if a large fluctuation occurs in external gas accretion, or as a result of a merger. However, conceptually gravitational quenching in combination with efficient feedback may provide a powerful process that could shut down global disk instability secularly, from the inside out (Martig et al. 2009). Indeed, recent simulations and semi-analytic models confirm that this process may play an important role in stabilizing disks, especially at late times.

In this paper we take advantage of the unique, high quality SINS/zC-SINF sample of z~1.5-2.5 SFGs presented in Förster Schreiber et al. (2013, henceforth FS13), along with ancillary HST WFC3 near-infared imaging by Tacchella, Lang et al. (in prep.) of the majority of the same galaxies, in order to test for evidence of the gravitational shutdown process discussed above. The SINS/zC-SINF sample provides deep, adaptive optics assisted SINFONI/VLT integral field (IFU) spectroscopy (Eisenhauer et al. 2003, Bonnet et al. 2005) of 35 z=1.5-2.5 SFGs. With these data it is now possible, for the first time, to derive significant constraints on the radial and mass variation of the Q-parameter in a statistically meaningful sample of massive high-z SFGs. We adopt a ΛCDM cosmology with $\Omega_m$=0.27, $\Omega_b$=0.046 and $H_0$=70 km/s/Mpc (Komatsu et al. 2011), as well as a Chabrier (2003) initial stellar mass function (IMF).

# 2. Observations and Analysis

## *2.1 Galaxy sample*

We have selected our galaxies from the adaptive optics assisted Hα IFU sample of FS13 (see also Förster Schreiber et al. 2009, Mancini et al. 2011), which in turn is drawn from several color and/or magnitude selected, rest-frame optical/UV imaging samples, with ground-based optical spectroscopic redshift identifications. We refer to the above papers for all details on the observations, data reduction and spectral/spatial analysis. The 35 z~1.5-2.5 SFGs in FS13 are representative of the overall near-main sequence, rest-optical/UVselected starforming population over the stellar mass range $logM_*$=9.2…11.5 but are somewhat biased toward bluer and more actively star-forming objects, largely because of the necessary (rest-UV) spectroscopic redshifts and the need for relatively high Hα surface brightness at least over some parts of the galaxies for detailed AO IFU follow-up (FS13).

From these 35 AO data sets (with a typical angular resolution of FWHM ~0.2", and spectral resolution of 85 km/s in K-band and 120 km/s in H-band) we selected rotation dominated galaxies, with

- a smooth, continuous velocity gradient along the morphological major axis, with no abrupt velocity jumps in the outer parts of the galaxy that might be indicative of a (major) merger. In most cases the projected velocity along the major axis levels off to an asymptotic value in the outer parts of the galaxy, as expected for a flat outer rotation curve,
- a projected velocity dispersion distribution peaking on/near the kinematic center, in many cases also identical with the center/nucleus of the galaxy on the ancillary HST images, and

- a sufficiently large size and signal to noise ratio per pixel to constrain the radial velocity distribution for dynamical modeling.

These criteria are necessary requirements if the large scale velocity field of the galaxy is to be dominated by rotation. They may not be sufficient to screen against minor mergers, or out of equilibrium disks formed in the aftermath of a major merger (c.f. Robertson et al. 2006, Robertson & Bullock 2008). With these selections, our sample retains 19 of the 35 SFGs in FS13. Figure 1 shows the integrated Hα images of these SFGs in the $M_*$-SFR plane. Table 1 summarizes their salient properties.

### *2.2 Kinematic and Mass Modeling*

We discuss our kinematic analysis and modeling in Appendix A, and we refer the reader to this section for all details and results on the individual galaxies. Figures A1 through A19 show the data and modeling results for all 19 SFGs of our sample. Table 1 is a summary of the inferred basic parameters, in particular, the dynamical mass, and estimates of $\Sigma_{mol\ gas}$ and Q for the 'inner' (central 0.1-0.15" in radius) and 'outer' disk/ring regions in each galaxy.

# 3. Results

Figure 1 shows the integrated narrow-line Hα maps for all 19 SFGs discussed in this paper, and arranged in the stellar mass – star formation rate plane. The diagonal continuous and dotted, white lines mark the location of the z~2 main-sequence, as well as star formation rates 4 times above and below. The dotted lines thus approximately denote the scatter around the main-sequence (Noeske et al. 2007). The 19 SFGs cover quite well the overall main-sequence population over almost a factor of 100 in stellar mass, also reflecting the same modest bias toward above main-sequence galaxies, especially at lower masses, as in the overall SINS/zC-SINF survey (Forster Schreiber et al. 2009, 2013 Mancini et al. 2011).

## *3.1 More than half of the SFGs exhibit Hα rings*

Figure 2 compares the inferred major axis, molecular gas surface density distributions (proportional to the Hα surface brightness distribution through the KS-relation, see Appendix A) of all 19 galaxies. It is immediately obvious that a significant number of these distributions are not centrally peaked but exhibit a ring distribution in observed Hα light. More than half of our sample (12 of 19) require modeling with a ring component in Hα light, or have $\Sigma_{outer}/\Sigma_{inner}>0.9$. The ring fraction appears to be largest at the high mass end: 7 to 8 of the 10 most massive SFGs have rings. However, this conclusion should be taken with some caution. The fraction of rings for smaller galaxies (typically lower mass) may be underestimated because of our instrumental resolution. In addition the non-Gaussian AO PSF shape (with substantial wings on the seeing limited scale) will have the tendency to fill in a compact ring brightness distribution. An example is the central Hα compact disk in

zC400569, which can be modeled well as an exponential (as in A14). A close inspection of the major axis position – velocity distribution and of the Hα surface brightness distribution (upper and lower right panels in A14), however, suggests that the exponential disk has a small central hole.

Typically 10-25% of the integrated Hα emission of the z~1-2.5 SFGs comes from a handful of bright star forming clumps (e.g. Förster Schreiber et al. 2009, 2011b, Genzel et al. 2011). The presence of these star forming clumps necessarily affects the inferred brightness distributions and radial cuts shown in Figure 2, However, the distributions shown in Figure 2 represent averages along the major axis on either side of the center, and across ~0.25"-0.3" perpendicular to the major axis, such that the impact of individual clumps is modest. In no case is the inference of a 'ring' just the result of a single bright off-center clump (Figure 1).

We have noted the occurrence of prominent Hα rings in several of the massive SFGs in the current sample before (BX482, zC406690: Genzel et al. 2008, 2011). The present study shows that such rings are common in massive high-z star forming disks. Wuyts et al. (2013) have investigated 473 3D-HST galaxies between z=0.7-1.5, taking advantage that for this sample (Brammer et al. 2012) both Hα and stellar surface densities are available at HST resolution (~0.2"). Wuyts et al. find from stacked light distributions that towards higher galaxy masses there is a clear trend toward a depression in central Hα emission and equivalent width, in excellent agreement with our findings and putting our conclusions here on a firm statistical footing. Hα rings are also found in z~0 star forming disks (Comeron et al. 2010).

An immediate question is whether these central depressions in the Hα distributions are intrinsic or whether they might be caused by differential extinction in flat or even centrally peaked intrinsic surface brightness distributions. The differential extinction hypothesis may be supported by the fact that in our sample SFGs with rings have an average Hα surface brightness 0.5-0.6 dex lower than in the centrally peaked cases. However, the much less extinction sensitive, Hα equivalent width in the stacked light distribution of the 473 z=0.7-1.5 SFGs studied by Wuyts et al. (2013) exhibits a central depression of 0.3-0.4 dex relative to the surrounding disk as well. Wuyts et al. (2013) also consider differential extinction between the R-band stellar light and Hα emission and correct the Hα emission appropriately, in the spirit of Calzetti et al. (2000) but considering physical extinction models better reproducing the rest-UV and Hα data of their 3D-HST high-z SFG sample. Even after such a correction the central Hα equivalent width depressions in the stacked light distribution remain, albeit at a smaller amplitude of 0.2-0.25 dex relative to the surrounding disk.

The work of Wuyts et al. (2013) suggests that differential extinction gradients are probably present and need to be taken into account but likely do not account for the frequent occurrence of Hα rings. The rings are probably an intrinsic property of the star forming gas.

## *3.2 Q-distributions are centrally peaked*

Figure 3 compares the inferred major axis Q-cuts for the 19 galaxies. In contrast to the observed Hα distributions and inferred molecular gas surface density distributions, the Q-distributions in all of our 19 SFGs are centrally peaked. With modest extrapolation to the spatial scales below the HWHM resolution (grey shaded region in Figure 3), 13 of the 19 SFGs exhibit $Q_{inner} \geq 1.3 \sim Q_{crit}$(thick disk, $f_{gas} \sim 0.5$). If the sample is divided in two by dynamical mass, the fraction of galaxies with $Q_{inner} > 1.3$ is the same in the two halves, but the average in the upper mass half has $<Q_{inner}> = 4$, significantly above the critical value, while the lower half has $<Q_{inner}>$

=1.3. Given our analysis and calibrations, this suggests that in these cases the nuclear regions must be globally stable to gravitational fragmentation. All but one of the 19 SFGs have Q significantly below unity in the outer parts and ring regions, fully consistent with the global/violent disk instability scenario, as shown previously by Genzel et al. (2011) for a subset of four of our SFGs.

The Toomre parameter is inversely proportional to molecular gas surface density, so naturally the question arises whether the centrally peaked Q-distributions are merely the consequence (and an artefact) of the central minimum in the observed Hα distributions. This question is explored in Figure 4, where we show again in the left panel the pixel by pixel Q-distributions of all 19 SFGs, as in Figure 3, with the κ ( R ) distributions obtained from the kinematic models, and the obvious strong trend of negative radial Q-gradients. To explore the dependence of the Q-gradients on gas surface density and κ-distributions independently, we replaced in the central panel of Figure 4 the κ distributions by a single average value for each galaxy. Now the gradients disappear for most points. Again with modest extrapolation to the radial scales below our resolution, $Q_{inner}$ remains greater than unity for much of the high mass half of the SFGs, but so does $Q_{outer}$. If so one would have to doubt the calibration of the Q-values, since obviously strong star formation does occur throughout the outer rings structures of these massive galaxies.

Finally in the right panel we let κ vary with R, as in the left panel, but now use a single value of $\Sigma_{mol\ gas}$ for each galaxy. While the outer Q-values are now somewhat higher, the inner values and especially the radial trends are pretty much the same as in the left panel. Figure 4 thus shows that it is the radial variations in κ, and not in $\Sigma_{mol\ gas}$ that largely drive the strong central Q-peaks in the massive half of the population. The κ distributions in many of our SFGs increase strongly toward the center, κ~1/R, because of the fairly flat or even inward raising rotation curves to 2-3 kpc (central upper panels in A1-19).

We conclude that the centrally peaked Q-distributions are influenced by but not dominated by the Hα ring distributions and are mainly driven by central mass concentrations increasing the central shear in the rotation curves.

### *3.3 Rings, central Q-peaks and inside-out quenching*

Assuming now that the inferred $\Sigma_{mol\,gas}$ and Q-distributions are a fair representation of reality, Figure 6 explicitly shows the dependence of Hα- and Q-distributions on galaxy (dynamical) mass. This Figure summarizes and strengthens the main results touched on before. The high-z disks in our sample, at all masses, are gravitationally globally unstable in most of their outer parts. The lower mass disks are also near the critical Q-value in their inner parts, consistent with their largely flat or even centrally peaked star formation distributions. However, above log $M_{dyn}$~10.8, strong mass concentrations inferred from the kinematics and rings in Hα drive the central Toomre parameters above unity in more than half of the galaxies.

The correlation between the presence of star forming rings and high central Q-values is strong. Of the 10 rings with $\Sigma_{molgas}$(inner) / $\Sigma_{molgas}$(outer)>0.9, 9 have $Q_{inner}$>1.3, and of the 13 galaxies with $Q_{inner}$>1.3 9 are rings. The ring size correlates with dynamical mass. The average ring radius for the lower mass half of our galaxies is 3.2 pc, while it is 5.6 kpc for the upper half.

One of our galaxies, GK2540, is an interesting special case. This system has relatively low mass (logM$_*$=10.3), with little evidence for a prominent central stellar mass concentration (Kurk et al. 2013). Its location below the main sequence means that this galaxy has less gas than the average galaxy at that mass (Magdis et al. 2012, Tacconi et al. 2013). GK2540 exhibits very low star formation and gas column densities (Figure 2), with Q barely dropping to unity in a very large, narrow star forming ring. GK2540 thus may be a case where the lack of star formation throughout the disk is largely driven by the lack of gas, perhaps as the result of currently low accretion, driving the galaxy below the main sequence line.

In summary, the data in the 19 rotation dominated SFGs studied in this paper are in excellent agreement with the hypothesis presented in section 1 that the global gravitational instability over time is suppressed from the inside out, as the galaxies grow in mass, shutting down global gravitational collapse, cloud formation and plausibly star formation over an increasing area of the most massive disk galaxies. Given that we see $Q_{inner} > Q_{crit}$ in about half of our massive SFGs, the gravitational quenching mechanism has to be quite efficient and have a high duty cycle.

An obvious next question is whether the galaxy-wide star formation rate in the Q-excess/ring galaxies is actually suppressed below that expected from the cold gas reservoir? For a clean test one would need direct estimates of the molecular gas masses of our sample for determining the gas depletion time scales (e.g. from CO observations, cf. Tacconi et al. 2010, 2013, Daddi et al. 2010, Genzel et al. 2010). Such data are unfortunately not currently available. An indirect hint comes from the fact that the main-sequence relation between stellar mass and star formation rate does not have a constant slope but flattens at high stellar mass, at all redshifts between ~0 and 2.5 (Whitaker et al. 2012). The ratio of specific star formation rates at $logM_*=10$ to $logM_*=11$ (in the regime where most of our rings are) is ~2, and increasing from high to low redshift. This drop indicates that the higher mass galaxies on average have lower molecular gas fractions, or indeed form stars less efficiently, than the lower mass galaxies at the same redshift. This difference can also be seen for our own sample when comparing the location of the galaxies relative to the slope 1 main sequence line in Figure 1.

### *3.4 What is the nature of the central mass concentrations?*

What is the nature of the central mass concentrations inferred from our dynamical modeling? In Figure 6 we compare the κ-values inferred from the modeling at the center on the horizontal axis, with those estimated from the inferred central stellar mass (filled blue circles, Tacchella, Lang et al. 2013) and molecular mass (open red squares) surface densities, as well as their sums (filled black squares) on the vertical axis, for the 13 galaxies where both can be estimated. Here we extrapolated the data and modeling inward to a fiducial radius of 0.4 kpc but the choice of a larger radius does not change the result. Given the substantial systematic uncertainties, the data for 11 of the 13 galaxies are in very good agreement with the hypothesis that the mass concentration inferred from our dynamical modeling is the same as the sum of cold (star forming) gas (inferred from the Hα brightness distribution) and stars (as estimated from the HST data). The ionized gas contributes only about 3-10% of gas mass (Genzel et al. 2011). It is possible that there is an additional substantial contribution from atomic hydrogen but at the typical column densities and pressures inferred from the molecular column densities most of the cold gas should be in molecular form (Blitz & Rosolowsky 2006).

In the galaxies with low $\kappa_{model}$ (largely identical with the galaxies with low dynamical masses), the central mass is dominated by gas. For the higher $\kappa_{model}$ galaxies (mostly higher mass), the fraction of stellar mass contributing to the central mass concentration becomes dominant. As we have seen in the last section, large central κ-values are the main drivers for the super-critical Q-values. Figure 6 suggests that the large central κ-values in turn are driven by the emergence of massive stellar bulges.

There are two outliers (BX482 and zC406690), where the dynamical modeling suggests the presence of much more mass than can be explained by either stars or molecular gas. These galaxies show very prominent Hα and stellar rings with little emission coming from the center, yet the kinematics indicates a major central mass concentration (Figures A13 & A16). One would have to resort to postulating either a concentration of sterile, non-star forming gas there, or very large nuclear extinction, or a combination of both. However, Tacconi et al. (2013) have reported direct CO 3-2 observations for both galaxies, which yield no or faint CO emission. Assuming a Galactic conversion factor, the faintness of the millimeter line emission is even inconsistent with the KS-estimate from Hα used in this paper, and certainly would not suggest extra gas (and dust). Given the low metallicity of both systems, it is possible that in these two cases that much of the molecular gas is 'CO-dark' due to UV photodissociation (Genzel et al. 2012). These 'dark' rings are currently not understood.

### *3.5 Caveats and alternatives*

As pointed out in the earlier sections, the conclusions in this paper, in addition to relying on a relatively small statistical sample, rest on a number of assumptions, all of which are uncertain or might be challenged,

1. the extinction correction of the Hα surface brightness maps relies on a uniform foreground screen model across each galaxy with extra attenuation towards HII regions relative to stars as proposed by Calzetti et al. (2000). This assumption (Calzetti et al. 2000, Calzetti 2001) does work empirically remarkably well even in very extreme, dusty local starburst regions in the local Universe, including ultra-luminous infrared galaxies (ULIRGs : Calzetti et al. 2000, Calzetti 2001, Engel et al. 2010, 2011). Yet it is doubtful that it also applicable to spatially resolved data (Genzel et al. 2013, Nordon et al. 2013, Wuyts et al. 2013). Moreover, the assumption of constant extinction across galaxies, even on resolved scales of ~ 1 – 2 kpc, is unrealistic. Local starburst galaxies, for instance, typically have extinctions peaking in the nuclear regions. However, the analysis of Wuyts et al. (2013) strongly suggests that radial trends in the H$\alpha$ vs stellar light/mass distributions are unlikely to be entirely caused by radial variations in extinction;

2. the empirical near-linear 'molecular KS-relation' that appears to hold on galaxy integrated and large scales in local and z~1-2 main-sequence SFGs (Bigiel et al 2008, Leroy et al. 2008, 2013, Genzel et al. 2010, Saintonge et al. 2012, Tacconi et al.2013, Daddi et al. 2010b), might break down on sub-galactic scales, in part because of the issue of extinction correction above (Genzel et al. 2013), and in part because of sampling and evolutionary effects (Onodera et al. 2010, Schruba et al. 2011, Calzetti, Liu & Koda 2012). Fortunately, points 1) and 2) to some extent counteract each other in the analysis of the current data;

3. the assumption of a constant local velocity dispersion in our modeling may be too simplistic, although the best current empirical evidence at both low and high-z is in support of just such a constant dispersion 'floor' (Heyer & Brunt 2004, Genzel et al. 2011, Davies et al. 2011, FS13, but see Green et al. 2010, Swinbank et al. 2012, Wisnioski et al. 2012). Specifically relevant to our study is the work of Genzel et al. (2011) and FS13 who searched for variations in $\sigma_0$ towards bright star forming clumps in $z\sim2$ SFGs in residual velocity dispersion maps, after correction for beam smeared rotation. They did not find any significant variations with local star formation surface density, with the possible exception of some nuclear regions, where the velocity dispersions appear to increase, most likely because of poorly modeled and unresolved nuclear motions. If these increases of velocity dispersion in the central regions were real and intrinsic, however, this would thus further increase Q and strengthen the results discussed above;

4. our kinematic/mass modeling delivers plausible but not unique model parameters, and rely on the assumption of equilibrium kinematics, which may not be justified in some cases. For instance, polar mergers may cause collisional ring galaxies (c.f. D'Onghia, Mapelli & Moore 2008). In fact one of our two 'dark centered' rings above the main sequence, BX482, has a nearby smaller companion about 3" to the south-east, and redshifted by about 750 km/s relative to the main galaxy. The companion is a compact star forming galaxy that is bright in H$\alpha$ (and CO, Tacconi et al. 2013). It is possible that in this case, the ring structure is a non-equilibrium result driven by a galaxy collision;

5. if the molecular gas depletion time scale were not constant but proportional to the local dynamical time scale, ring structures may naturally form as a result of this radial dependence, rather than from gravitational quenching. Future high resolution molecular observations of our SFGs will be able to test such a hypothesis. There is no dependence of the depletion time scale on galactic radius in z~0 star forming disks (Leroy et al. 2008, 2013).

### *3.6 Comparison to low-z disk galaxies*

In contrast to the situation discussed here for high-z star forming disks, recent observations of massive ($logM_*>10$) z~0 SFGs suggest that the Toomre parameter does not play a major role in controlling galactic star formation. In the HERACLES CO 2-1 survey at the IRAM 30m telescope (in combination with GALEX UV data, SINGS/Spitzer 24µm data and THINGS HI data) Leroy et al. (2008) have carried out spatially resolved (400-800 pc resolution) mapping of the gas – star formation relation in 12 massive spirals ($logM_*=10.1$-$10.9$) and 11 dwarfs ($logM_*=7.1$-$9.9$). From these data Leroy et al. construct the radial dependence of the Q-parameter (in gas as well as gas + stars). Their Figure 9 (equivalent to our Figure 3) does not show any strong trends of $Q_{gas}$ or $Q_{gas+*}$ with galacto-centric radius. The average massive spiral at z~0 has Q~2-4 throughout its disk and nuclear regions and thus is stable against gravitational fragmentation. The galactic gas depletion time scale (the inverse of the "star formation efficiency") does not vary with Q.

A particularly instructive case is the grand design spiral M51 (NGC5194), which has become a benchmark system for studying star formation on galactic scales. Hitschfeld et al. (2009) show that $Q_{gas}$ and $Q_{gas+*}$ on average range between 2 and 4 throughout the disk of M51, but dip to values near or even slightly below 1 on the spiral arms in the outer disk. However, the gas depletion time scale in these arms does not differ from the interarm regions; strong spiral arms may have $Q \leq Q_{crit}$ but do not result in more efficient star formation (Foyle et al. 2010). Elmegreen (2011) concludes that "the primary effect of a spiral is to concentrate the gas in the arms without changing the star formation rate per unit gas". In the analysis of the first CO 1-0 IRAM PdBI observations of M51 within the PAWS high resolution program Meidt et al. (2013) even conclude that in those parts of the spiral arms with strong streaming motions and large pressure gradients, GMCs may actually be driven to lower star formation efficiency. Spiral arms may thus act on the one hand to collect and form GMCs and on the other also decrease their star formation efficiency.

We suspect that it is the strong difference in molecular gas fractions that drives the difference in galactic gas fragmentation and star formation in the regimes of high-z and local disk galaxies.

## *3.6 Comparison to theoretical expectations*

As we have discussed in section 1 the occurrence of star forming rings in galaxies with high central Q values is a natural outcome of quenching of radial gas transport into the inner disk regions. As Q is below unity in the outer regions, gravitational torques and clump-clump interactions will lead to angular momentum redistribution, driving angular momentum outwards and gas inwards. If the in-spiraling material is gas-rich, that is, if the star formation time scale is longer than the in-fall timescale (Dekel & Burkert 2013), the gas will reach the inner region where the disk is stable due to $Q > 1$ and where radial transport is suppressed. At the boundary between the gravitationally stable inner region and the unstable outer region the in-falling gas will accumulate, generating a gas-rich ring with enhanced star formation. Star forming rings driven by the combined effect of gravitational instability and radial gas transport indeed occur frequently in recent cosmological galaxy formation hydro-simulations with sufficient resolution to study sub-galactic scales, and are more common in more massive systems (Ceverino et al. 2010, Genel et al. 2012, Ceverino, priv.comm.).

What happens to the 'sterile' gas collecting in the inner regions? Will it not accumulate there until Q drops again sufficiently to rekindle the instability? In the theoretical studies of these processes the radial transport becomes inefficient at the same time as the gravitational instability stops, drastically decreasing the matter transport into the center (Martig et al. 2009, Ceverino et al. 2010, Cacciato, Genel & Dekel 2012, Forbes et al. 2013). During that phase star formation continues in the central regions at a lower rate. Thus there may be little accumulation. Alternatively AGN feedback may efficiently eject gas that is transported into the nuclear regions.

Why do especially the massive galaxies have large bulge masses with star-formation-quenched inner regions and rings? It is tempting to identify these galaxies as being in their last active phase of star formation. Gas in their inner regions has already been depleted by star formation with refueling through radial inflow from the outer, gas-rich disk regions being suppressed as discussed above. The fact that most of the massive rings in Figure 1(with the exception of the 'dark rings' BX482 and zC406690, see 3.4) are somewhat below the main sequence line may suggest that also gas refueling by infall from the cosmic web has slowed down and that these galaxies are in the process leaving the main sequence with their star formation rate decreasing. Adopting $SFR = M_{molgas}/t_{depl}$, the gas mass in this final phase is expected to decrease exponentially with an e-folding time scale of $t_{depl} \sim 1$ Gyr. A change in the star formation rate by 0.3 dex (as in Figure 1) then corresponds to an evolutionary timescale comparable to the depletion time scale, which would appear reasonable.

# 4. Conclusions

We have presented high quality adaptive optics assisted SINFONI/VLT integral field spectroscopy of Hα line emission and kinematics in 19 rotation dominated near-main sequence star forming galaxies, ranging in stellar mass from $4 \times 10^9$ to $3 \times 10^{11}$ $M_\odot$.

We have used the high quality kinematic information in these data sets to deduce the radial dependence of circular velocity, dynamical mass and epicyclic frequency, as well as the local velocity dispersion in the outer parts of these galaxies. We have taken the Hα surface brightness distributions, corrected globally for extinction, together with the z~2 PHIBSS calibration of the molecular Kennicutt-Schmidt relation (Tacconi et al. 2013), to construct molecular column density maps. Combining the kinematic modeling and Hα mapping we were then able to derive major-axis cuts of the Toomre Q-parameter for all 19 SFGs in our sample.

We find that in all of our galaxies Q decreases from inside out, where it is substantially below unity. All outer disks thus are globally unstable to gravitational fragmentation. In contrast the Q value near the center, $Q_{inner}$, increases above the critical value of about 1.3 for half to two thirds of our sample. At the same time a similar fraction of our galaxies exhibit Hα rings, rather than centrally peaked, Hα distributions. The probability to both show a ring structure and $Q_{inner} \geq Q_{crit}$ is strongly correlated and increases with dynamical mass. The presence of rings and super-critical Q values is correlated with the emergence of massive central stellar bulges, and a drop in the specific star formation rate. Keeping in mind the possible pitfalls and uncertainties in our analysis (un-modeled extinction gradients, radial variations in velocity dispersion, and departures from linearity in the relationship between star formation and molecular gas surface density etc.), our findings are in plausible agreement with an efficient inside-out, low-to-high mass suppression/reduction of the gravitational instability in z~2 SFGs that has been predicted by several recent theoretical papers.

We find that the super-critical central Q values are mostly driven by the presence of a central mass concentration driving up the central shear. In 11 of the 13 SFGs in our sample with HST WFC3 imagery, the mass concentrations inferred from our modeling are consistent with the sum of the molecular gas and stellar mass near the centers. The central molecular mass concentrations dominate for the low dynamical mass galaxies of our sample, while the stellar contribution becomes significant and even dominant in most of the high mass systems. This finding is consistent with the current theoretical picture that gas and newly formed stars in the gas-rich high-z disks are efficiently driven inward by torques and dynamical friction and establish a fast growing star-forming bulge there.

The gravitational quenching process discussed above is unlikely to lead by itself to the long-term quenching of star formation but probably requires the participation of other players, such as the decrease of gas accretion rates with halo mass and cosmic time, and the removal of non-star forming gas by feedback processes, such as AGN driven nuclear winds.

*Acknowledgements: We thank the staff of Paranal Observatory for their support. We are grateful to Daniel Ceverino for comments on the paper and for communicating to us the frequency of star forming rings in his hydro-simulations. We thank Adam Leroy, Eva Schinnerer, Andreas Schruba and Fabian Walter for comments on the Q-parameter in z~0 disks.*

# Appendix A. Kinematic modeling of the individual galaxies

As discussed previously in Genzel et al. (2011, to which we refer for details), our kinematic analysis and modeling incorporate the following steps,

1. *spectral extraction*: we extracted spectra along the structural/kinematic major axis using a synthetic slit with an effective sampling of 0.1" to 0.2" along the slit, and a width of 0.25" to 0.3" perpendicular to the slit. Gaussian fits deliver Hα surface brightness I, projected velocity v and projected velocity dispersion σ for each pixel, along with their fit errors;

2. *disk modeling*: we constructed rotating disk models fitting the observational constraints I(p), v(p) and σ(p) as a function of projected major axis position offset p, from the kinematic/stellar centroid of the galaxy. These disk models compute data cubes from input structural parameters (c.f. Cresci et al. 2009). The main parameters are the disk's center position, its inclination and major axis orientation on the sky, as well as its mass and light distributions as a function of radius, its total dynamical mass and a constant additional velocity dispersion assumed to be isotropic. Position angle, inclination and centroid are determined from the morphology of the Hα and (where possible) HST images, and (for the centroid) from the zero crossing of the observed rotation curve, assuming reflection symmetry in velocity along the major axis. The model data are then convolved with the angular and spectral resolution instrumental profiles and sampled at the observed pixel scale. Surface brightness, velocity and velocity dispersion cuts along the major axis are then extracted as for the data. The total dynamical mass $M_{dyn}$, and the light and mass distributions (not necessarily identical) are then varied to achieve a fit to the data along the major axis. We have also carried out 2D fitting of I, v and σ but find that the major axis information captures the essential information needed for the mass modeling.

In all cases we start with the assumption of an exponential distribution in both mass and Hα light, with a half-mass/light radius taken from the analysis in FS13, based on 2D Sersic fitting and a curve of growth analysis of the integrated Hα flux distribution. In more than half of the sample an exponential is obviously not a good fit to the Hα light distribution (Table 1, Figures A1- A19). A Gaussian ring, or a ring plus an exponential are then adopted to better match the surface brightness cuts. The average ring radius in Table 1 is 4.5 kpc but there is a large scatter from 1.5 to 9 kpc. In 5 of the 19 cases, the steepness of the central major axis velocity gradient, combined with a prominent peak of velocity dispersion near the kinematic centroid requires a mass distribution more compact than an exponential, for instance the combination of the original $R_{1/2}$ exponential with an additional nuclear mass concentration, assumed for simplicity to be a Gaussian. The specific derived model components and parameters are not unique, nor necessarily well constrained. However, the velocity data DO robustly constrain the mass concentration and the rotation curve, and give an estimate of its amplitude within the central few kpc, relative to the overall disk. The primary outputs of this modeling are, first, the total dynamical mass within R<10-12 kpc (the radius range mapped by the Hα data); second, the intrinsic velocity dispersion (assumed to be constant and isotropic) required to match the observed velocity dispersion in the outer part of the disk (which is little or not affected by beam smeared rotation); and third, the intrinsic rotation curve, and thus the epicyclic frequency distribution κ ( R ), as introduced in section 1. The absolute values of the rotation velocity and dynamical mass depend linearly and in squares on the sine of the inclination, *i*. The inferred inclinations from the morphological aspect ratio of the Hα and or stellar distribution typically are uncertain to ±5 up to 20 degrees (e.g. Cresci et al. 2009). This implies uncertainties in velocity and mass of 20 to 50% for inclinations >50 degrees, but can lead in extreme cases to uncertainties of a factor of several for nearly face on systems. The inclination dependence also affects the overall value of the epicyclic frequency needed to determine Q, but not its radial distribution.

3. *molecular gas surface density distribution*: we used the Kennicutt-Schmidt (KS) relation to infer molecular gas surface densities from star formation surface densities. For calculating star formation rates from integrated Hα data we applied the conversion of Kennicutt (1998a,b) modified for a Chabrier (2003) IMF ($SFR$=L(Hα)$_0$/2.1x10$^{41}$ erg/s). We first corrected the observed Hα maps for broad emission that come from outflows (Newman et al. 2012, Förster Schreiber et al. 2013b). For this purpose we removed the large scale velocity shifts due to rotation pixel by pixel, and then computed an integrated 'narrow' line Hα map by rejecting Hα emission outside the narrow line core. This method does not, however, correct for the contribution of the broad emission within the narrow line core, which can be substantial in very bright clumps and in nuclear regions (Genzel et al. 2011, Förster Schreiber et al. 2013b). In those cases we attempted a more complete removal of the broad emission by two component fitting in each pixel. We then converted the integrated narrow line Hα map to a star formation surface density map from the Kennicutt (1998b) calibration above. We next corrected the observed star formation surface density map for spatially uniform extinction with a Calzetti (2001) extinction curve (A(Hα)=7.4 E(B-V)), including the extra 'nebular' correction (A$_{gas}$=A$_{stars}$/0.44) introduced by Calzetti (2001). We determined E(B-V) from the integrated UV/optical photometry of the galaxies. Förster Schreiber et al. (2009), Mancini et al. (2011) and Wuyts et al. (2011a) find that including the extra nebular correction brings Hα- and UV-continuum based star formation rates of z~2 SINS/zC-SINF galaxies into better agreement than without such a correction (but see Reddy et al. 2010, Kashino et al. 2013). However, the Calzetti modified screen approach probably breaks down for spatially resolved data (e.g. Genzel et al. 2013), since in reality the extinction is a combination of the large scale dust distribution in the diffuse interstellar medium, with local dust concentrations associated with the individual star forming clouds (Nordon et al. 2013, Wuyts et al. 2013). The integrated Calzetti screen approach taken by necessity in this paper (for lack of spatially resolved $A_V$-maps) probably underestimates molecular columns in the densest, dustiest star forming clumps and in nuclear gas concentrations.

To convert star formation surface densities obtained in this way to molecular gas surface densities, we used the PHIBSS calibration from Tacconi et al. (2013), based on galaxy integrated CO measurements in massive main-sequence SFGs between z~0 and 2.5. PHIBSS yields a simple linear KS relation and a slowly varying depletion time scale, $M_{mol\ gas}\ (M_\odot) = t_{depl}(z) \times SFR\ (M_\odot\ yr^{-1})$, with $t_{depl}=1.5\times10^9\ (1+z)^{-1}\ (yr)$ (see also Saintonge et al. 2011, 2012, 2013). While this calibration is probably fairly robust on galaxy integrated scales (Daddi et al. 2010a, Magdis et al. 2012, Magnelli et al. 2012), with a systematic uncertainty of ±0.3 dex because of uncertainties in the CO to molecular gas conversion factor, the spatially resolved molecular KS- relation may be steeper than linear, both in the local Universe and at high-z (Kennicutt et al. 2007, Daddi et al. 2010b, Heidermann et al. 2010, Kennicutt & Evans 2012, Genzel et al. 2013). For slope

N~1.3 proposed by Kennicutt et al. (2007) in M51 (N=$\log\Sigma_{SFR}/\log\Sigma_{mol\,gas}$), for instance, such a non-linear relation would have the tendency of lowering the inferred molecular gas columns in the brightest star formation regions (by 60% over a factor of 10 in surface density), plausibly counteracting some of the extinction effects discussed above.

4. ***Q-distribution***: we finally combined the information on $\sigma_0$ and $\kappa(R)$ from the kinematic modeling, with the gas distributions $\Sigma_{mol\,gas}$ from the Hα data to derive the Toomre parameter for each pixel along the major axis, using Equation 2. Uncertainties in Q are derived from the pixel by pixel uncertainties in $\Sigma$. The uncertainties in $\sigma_0$ and $\kappa$ are not included, as they mostly enter the larger systematic uncertainties but much less so the radial variations. Including these uncertainties would increase the average fractional error of Q from ~0.15 to ~0.4.

5. *stellar surface density distribution*: Tacchella, Lang et al. (2013) have analyzed the J- and H-band WFC3 images of 13 of the 19 SFGs discussed in this paper and inferred intrinsic stellar mass surface density maps. From these maps we extract the central values in the same apertures as for the 'inner' molecular gas surface densities (typically with a radius of 0.1-0.15"), to derive total inner (central) baryonic surface densities.

In Figures A1 through A19 we show for all 19 galaxies the I, v, and σ cuts extracted from the Hα data as described above, along with the fitted models and the inferred Q, $\Sigma_{mol\,gas}$ distributions. In Table 1 we summarize the inferred basic parameters, and in particular the dynamical mass and estimates of $\Sigma_{mol\,gas}$ and Q for the 'inner' (central 0.1-0.15" in radius) and 'outer' regions in each galaxy. The latter is typically an average over 0.2"-0.3", on either side of the nucleus and centered near $R_{1/2}$, or the ring maximum identified in the modeling.

**Figures**

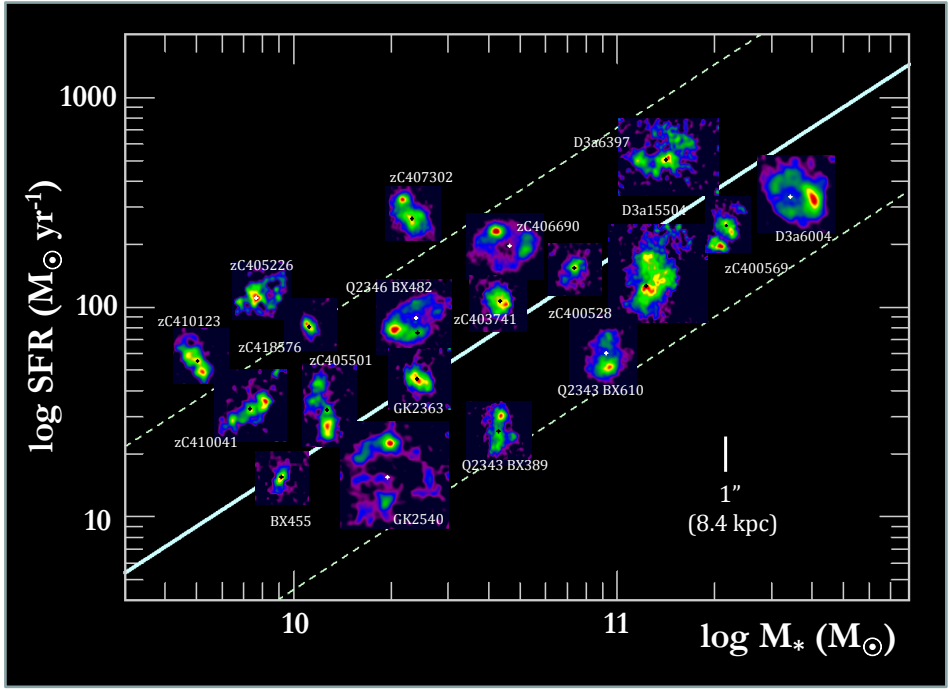

Figure 1. Integrated Hα maps of the 19 disks in this paper, in the stellar mass –star formation rate plane. The FWHM angular resolution of these maps is ~0.21-0.27", and all galaxies are on the same angular scale (the white vertical bar indicates 1" (~8.4 kpc)). The color scale of the brightness distributions is linear and auto-scaled. The continuous white line marks the location of the z~2 'main-sequence' with an assumed slope of 1 (sSFR=SFR/$M_*$=const, e.g. Daddi et al. 2007, Rodighiero et al. 2010, Whitaker et al. 2012), with the dashed lines denoting star formation rates ~4 time above and below the white line, roughly indicating the scatter of the star formation main sequence. Several of the images are rotated in order to better fit onto the plot.

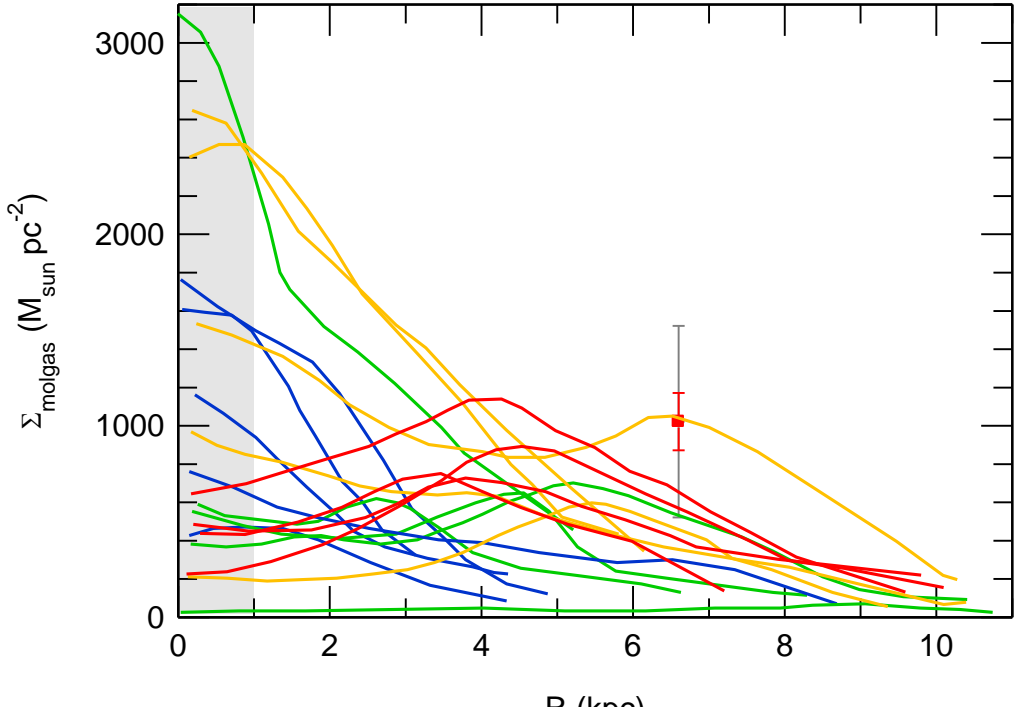

Figure 2. Inferred radial molecular gas surface density distributions (from the observed Hα brightness distributions at a typical FWHM resolution of 2 kpc, averaging the values on either side of the center) for all 19 SFGs in this paper, separated by dynamical mass in the lowest 5 (blue: $10.36 \leq \log M_{dyn} \leq 10.5$), next 5 (green: $10.68 \leq \log M_{dyn} \leq 10.93$), next 5 (orange: $11.04 \leq \log M_{dyn} \leq 11.28$) and highest bin (red: $11.34 \leq \log M_{dyn} \leq 11.41$). Typical statistical (red) and systematic (grey) uncertainties are indicated. The appearance of ring distributions, especially among the two highest mass is apparent. The bottom green curve is GK2540. The grey-shaded area on the left denotes the radius-regime that is below the average HWHM instrumental resolution, and thus represents a somewhat uncertain inward extrapolation.

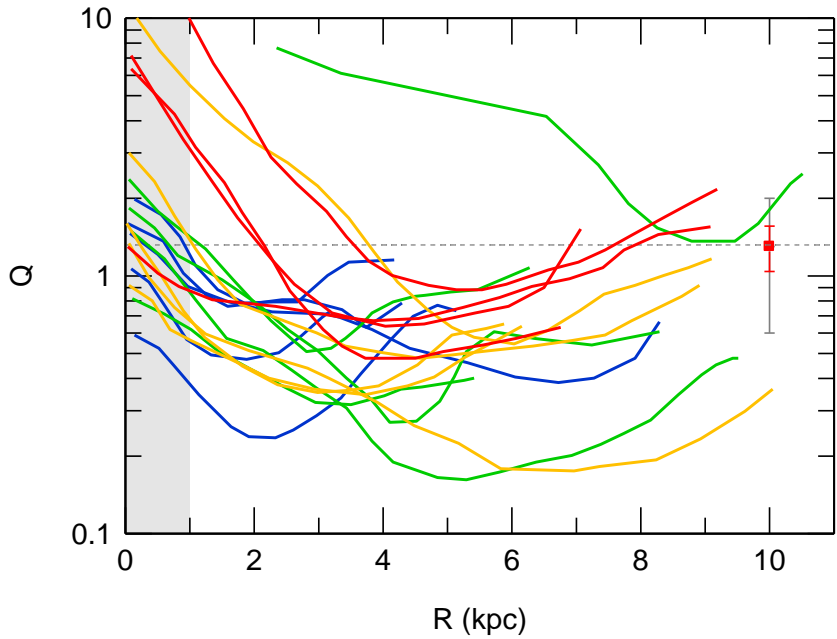

Figure 3. Radial distributions of the Toomre Q-parameter for the 19 SFGs in this paper, separated as in Figure 2 by dynamical mass in the lowest 5 (blue: $10.36 \leq \log M_{dyn} \leq 10.5$), next 5 (green: $10.68 \leq \log M_{dyn} \leq 10.93$), next 5 (orange: $11.04 \leq \log M_{dyn} \leq 11.28$) and highest bin (red: $11.34 \leq \log M_{dyn} \leq 11.41$). Typical statistical (red) and systematic (grey) uncertainties are indicated. The dashed horizontal line marks $Q_{crit}=1.3$, for a thick gas-rich disk with $f_{gas} \sim 0.5$. The grey-shaded area on the left denotes the radius-regime that is below the average HWHM instrumental resolution, and thus represents a somewhat uncertain inward extrapolation.

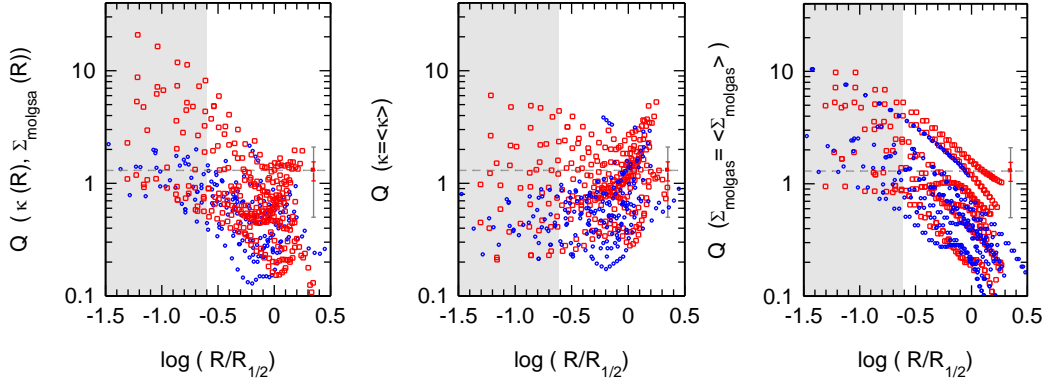

Figure 4. Distribution of Q-values for each pixel and all SFGs, separated in two mass bins (blue: 11 lowest mass, red: 8 highest mass). The left panel depicts the same data as in Figure 3, with $\Sigma_{mol\,gas}$ derived from the H$\alpha$ data, and $\kappa(R)$ and $\sigma_0$ derived from the dynamical modeling. The central panel again uses the same molecular surface densities and velocity dispersions as the right bin but instead applies a constant average $\langle\kappa\rangle$ value for each galaxy. The right panel instead uses $\kappa(R)$ and a constant (median) value for the molecular surface densities. A comparison of the three panels shows that the strong dichotomy of strongly gravitationally unstable (Q<1) gas in the outer disks and stable (Q>1.3) gas in the nuclear regions, especially for the more massive SFGs, is more driven by the radial variation in $\kappa$ than in $\Sigma_{mol\,gas}$. The red and grey error bar denote the typical statistical and systematic uncertainty of the data. The grey-shaded area in each panel denotes the radius-regime that is below the average

HWHM instrumental resolution, and thus represents a somewhat uncertain inward extrapolation.

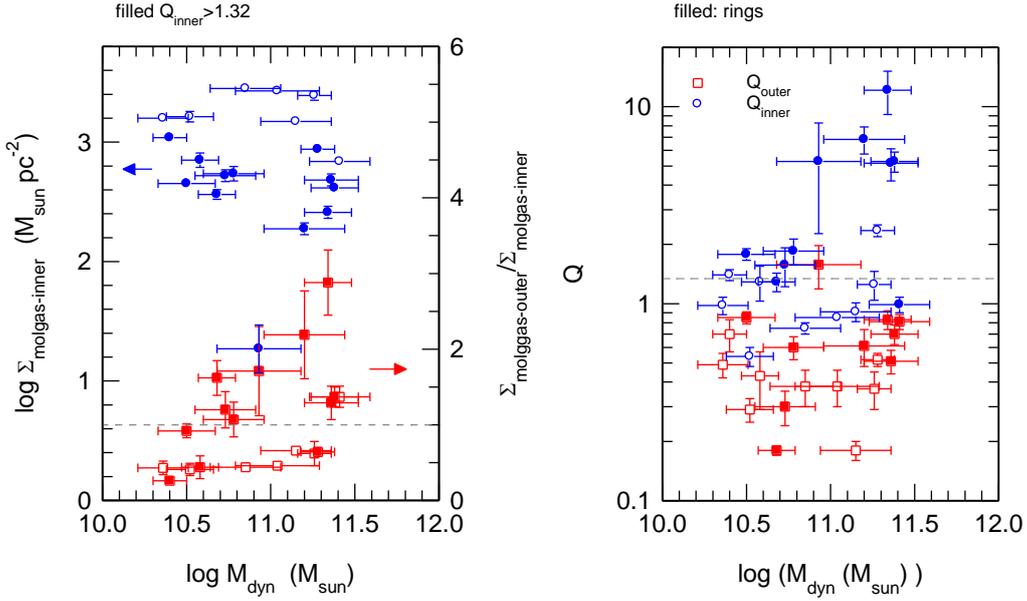

Figure 5. Evidence for radially quenching of gravitational fragmentation in z~1.5 – 2.5 disks. The left panel shows the central molecular gas surface densities (blue circles, left axis), and the ratio of average outer disk (near $R_{1/2}$, or the ring maximum) to central surface densities (red squares, right axis), as a function of dynamical mass. Galaxies with $Q_{inner} \geq 1.32 = Q_{crit}$(thick, $f_{gas}$~0.5) are denoted by filled symbols. The right panel shows the inner (blue circles) and outer (red squares) average values of the Q-parameter as a function of dynamical mass. Rings galaxies ($\Sigma_{molgas}$(inner) / $\Sigma_{molgas}$(outer)>0.9) are denoted by filled symbols.

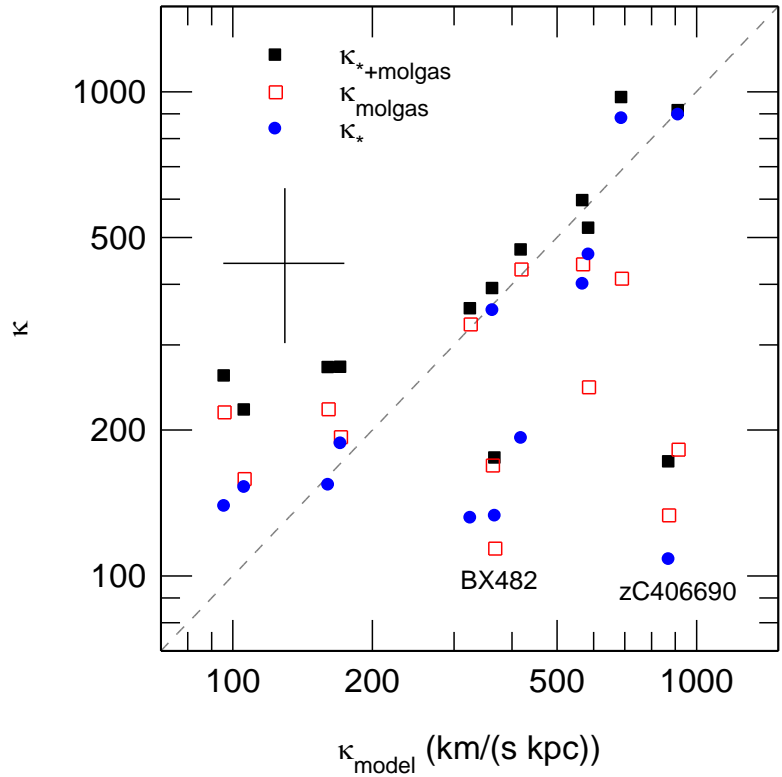

Figure 6. Comparison of the epicyclic frequency determined from the dynamical modeling (horizontal axis) with the epicyclic frequency determined from the observed central stellar mass surface density (filled blue circles), from the inferred molecular gas surface density (open red squares), as well as their sum (filled black squares) on the vertical scale. The fiducial radius at which this comparison is made is 0.05" (0.4 kpc). Given the estimated systematic uncertainties (large black cross), the combination of gas and stellar mass can plausibly account for the central mass inferred from the gas kinematics (the dashed grey line indicates a ratio of unity), with the exception of the two 'dark centered' galaxies BX482 and zC406690. The central shear is dominated by gas for the galaxies with low shear and with the exception of BX482 and zC406690, there is a tendency for the stellar component to become dominant for the higher $\kappa$ systems.

Figures A1-A19. Results of the dynamical modeling of the individual galaxies. The three bottom panels show the observed Hα surface brightness (left), velocity (middle) and velocity dispersion (right) distributions as blue filled circles, as a function of major axis offset (along the dotted white line in the upper right Hα, or Hα+continuum images). The typical software slit width perpendicular to the major axis is 0.25 to 0.3". As described in the text, we created simple rotating disks with one or two mass and Hα luminosity density components that fit these data. The surface density, circular velocity and dynamical mass distributions of these input models are shown as red continuous lines in the top row; in some of the cases the surface brightness models (dotted red lines in the upper left) differ from the mass distributions. The projection of these models onto the major axis software slits, smoothed to the spatial and spectral instrumental resolutions, are shown as red dotted curves in the lower three panels. The bottom right panel compares the distributions of the inferred molecular surface density distribution (red, right axis) and of the inferred Toomre Q-parameter (filled blue circles, left axis) along the kinematic major axis. The HST WfC3 J/H images used in the upper right panels are from Tacchella, Lang et al. in prep., in a few cases we also used the continuum from the SINFONI cubes themselves. The Figures are sorted from low to high dynamical mass, as in Table 1.

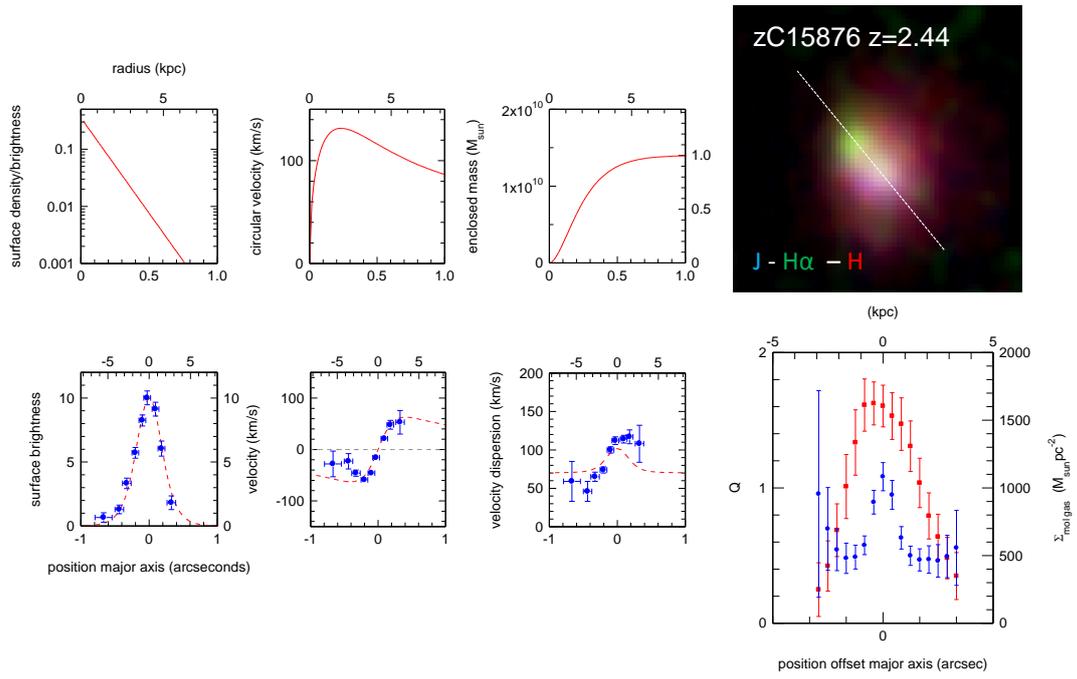

Figure A1.

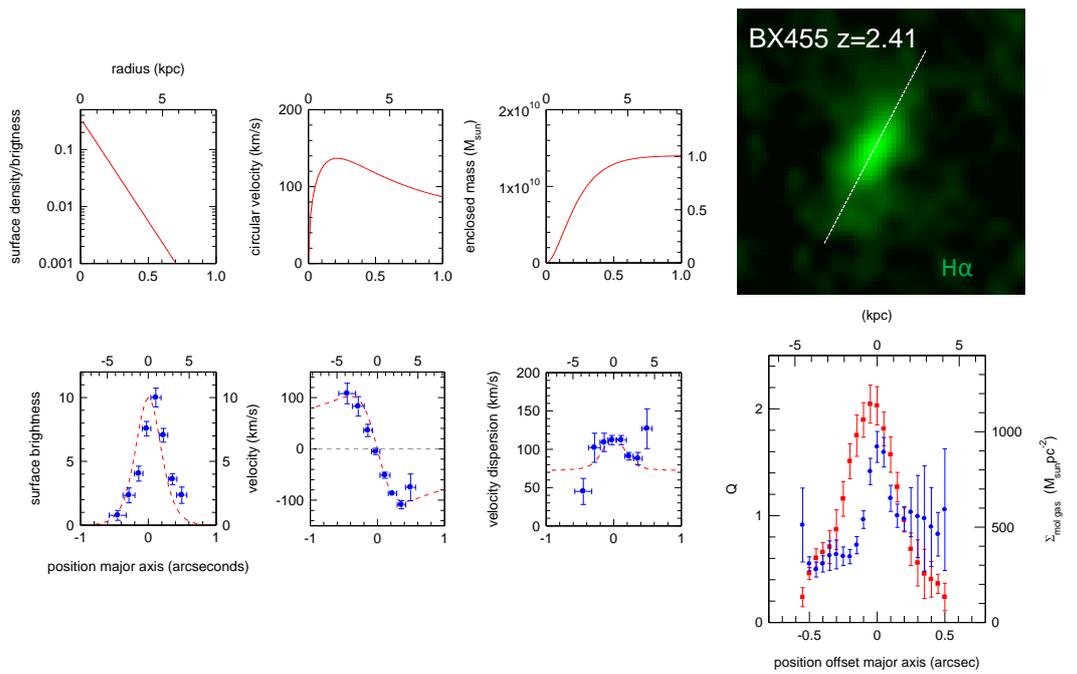

Figure A2.

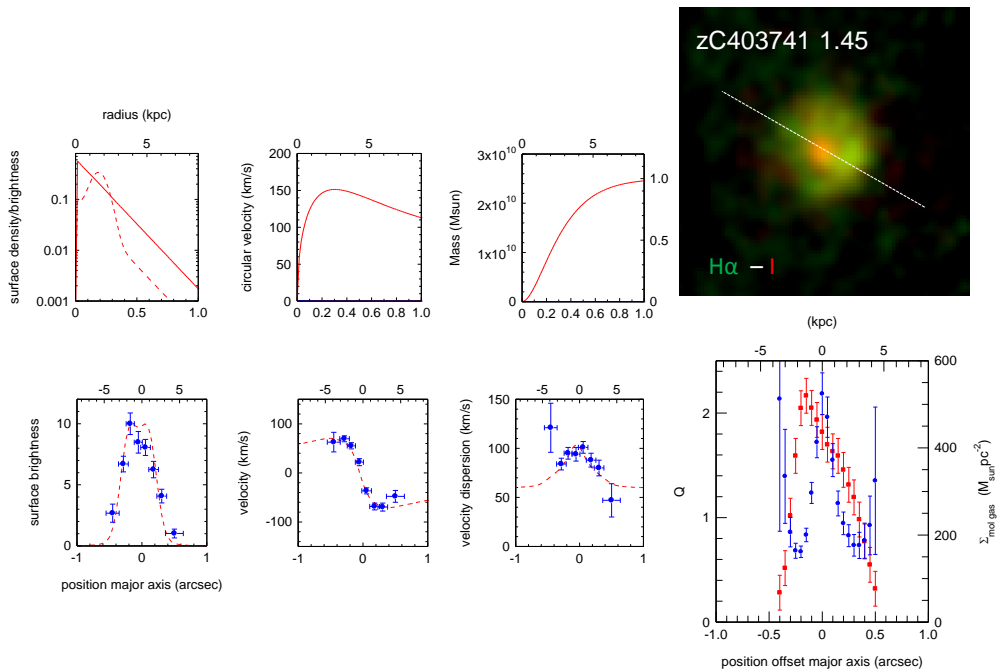

Figure A3

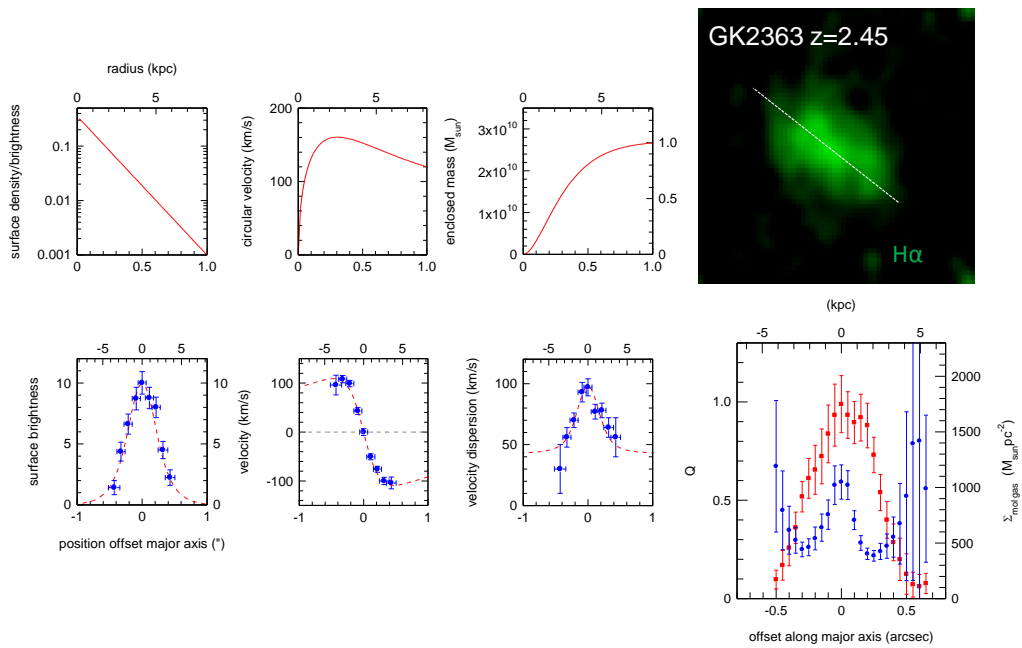

Figure A4.

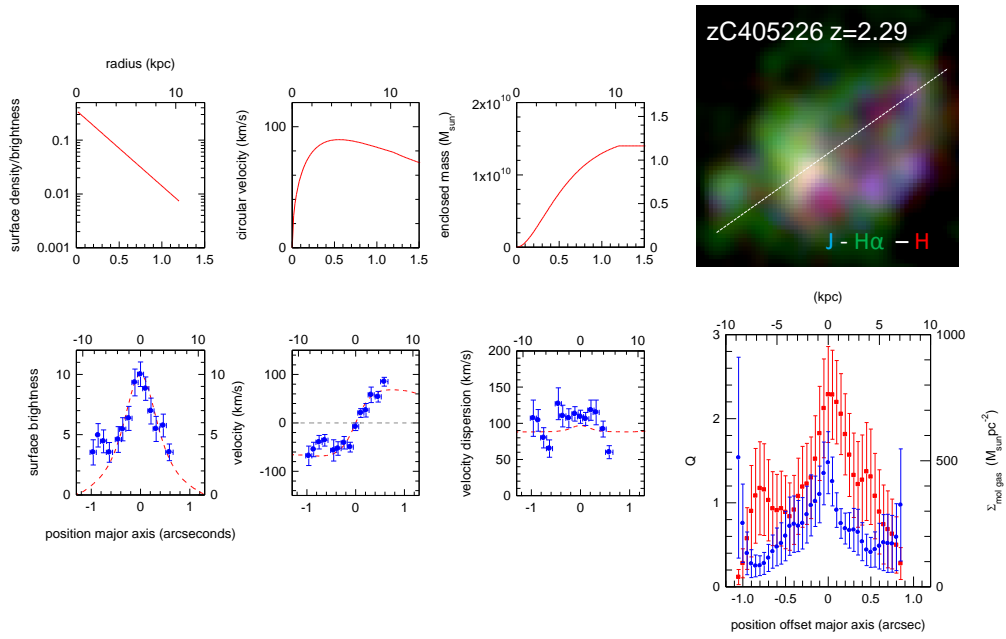

Figure A5.

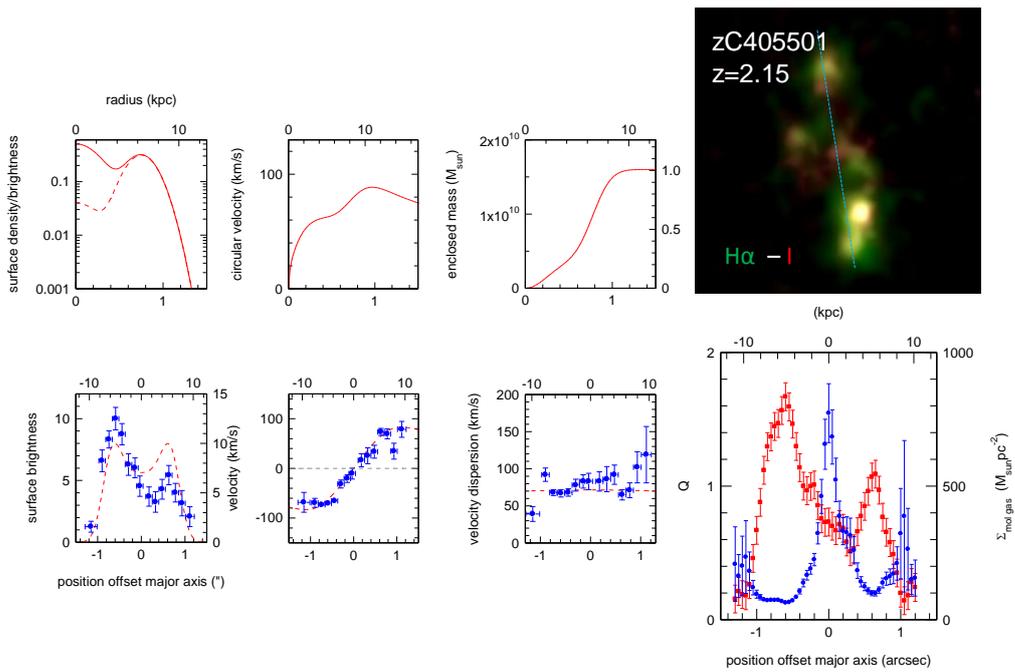

Figure A6.

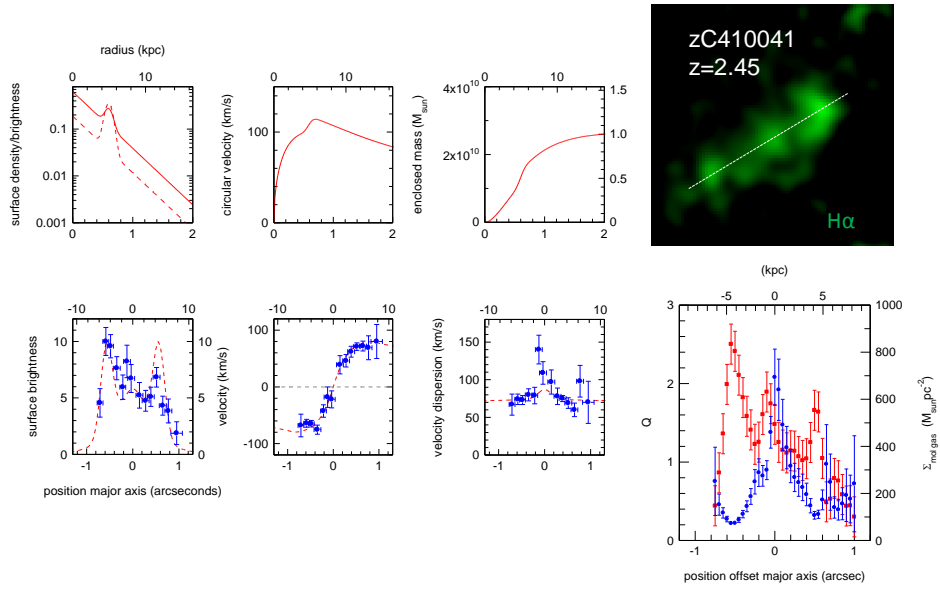

Figure A7

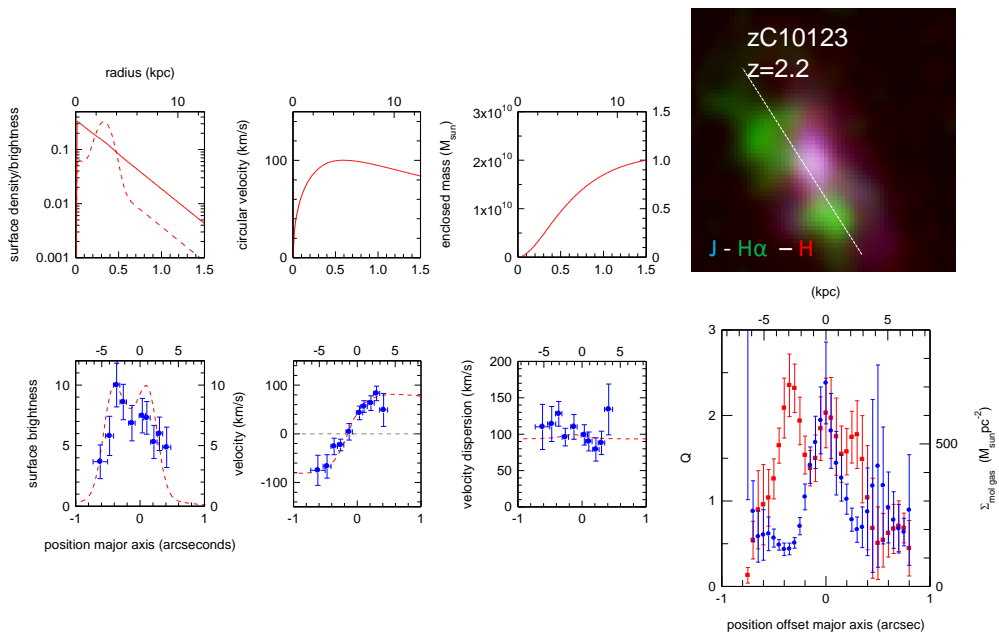

Figure A8.

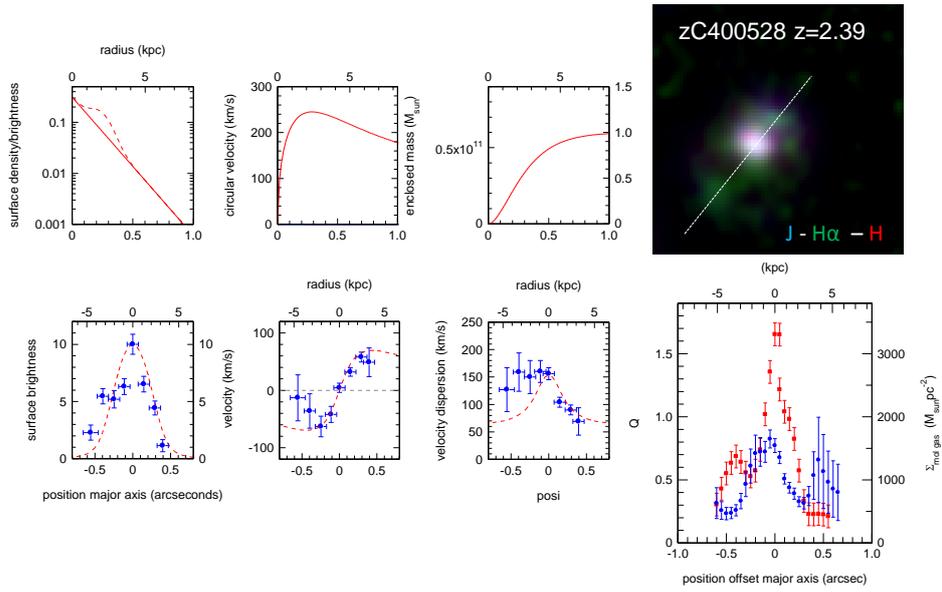

Figure A9.

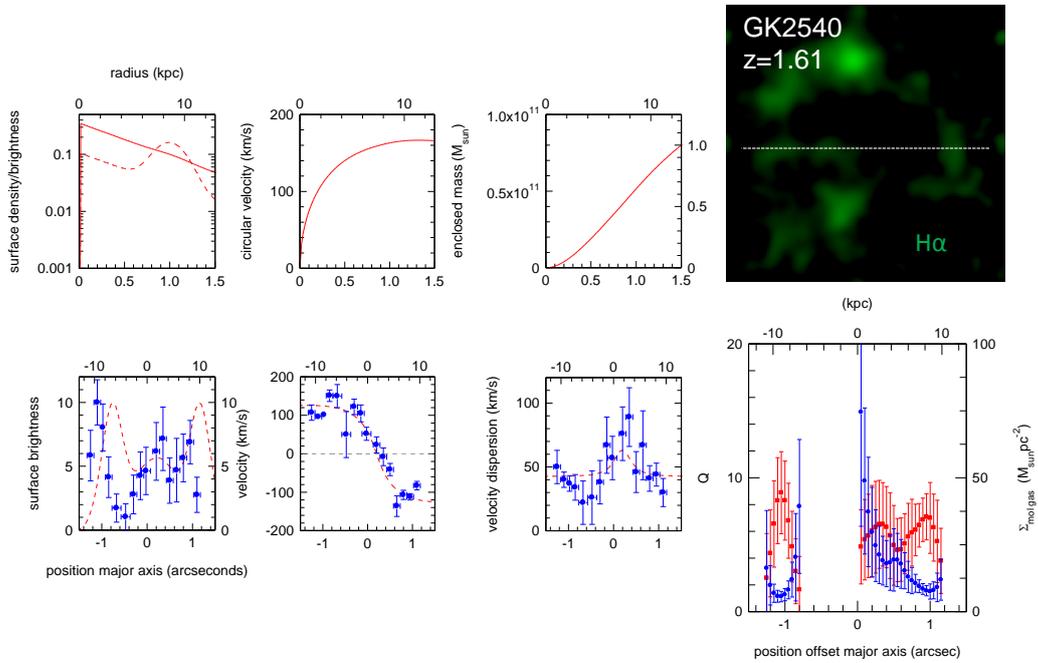

Figure A10.

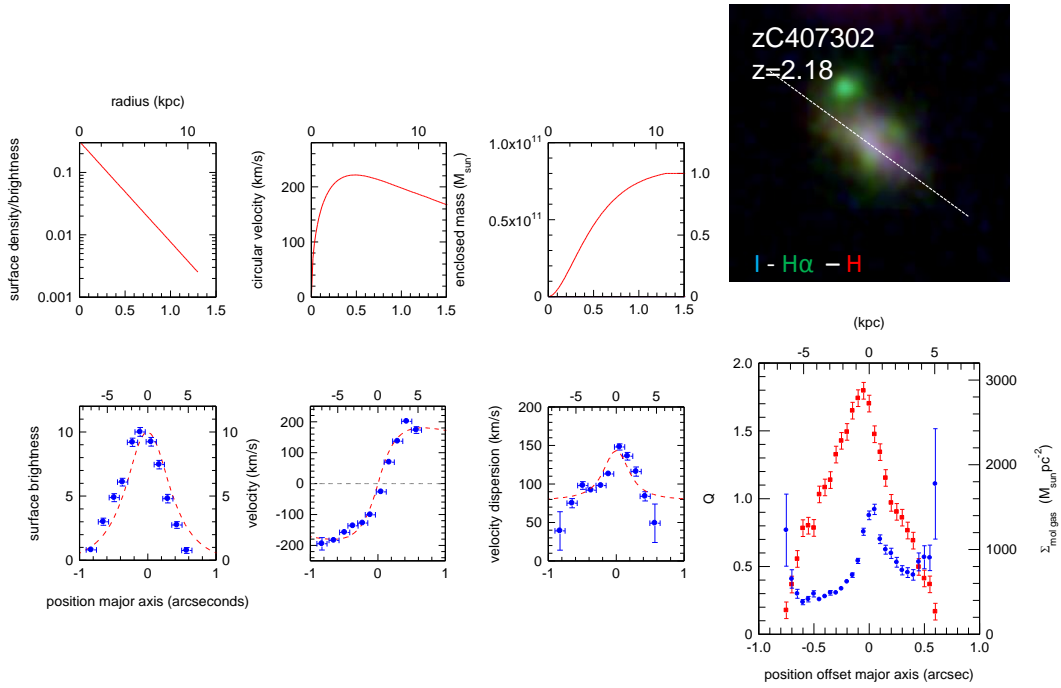

Figure A11.

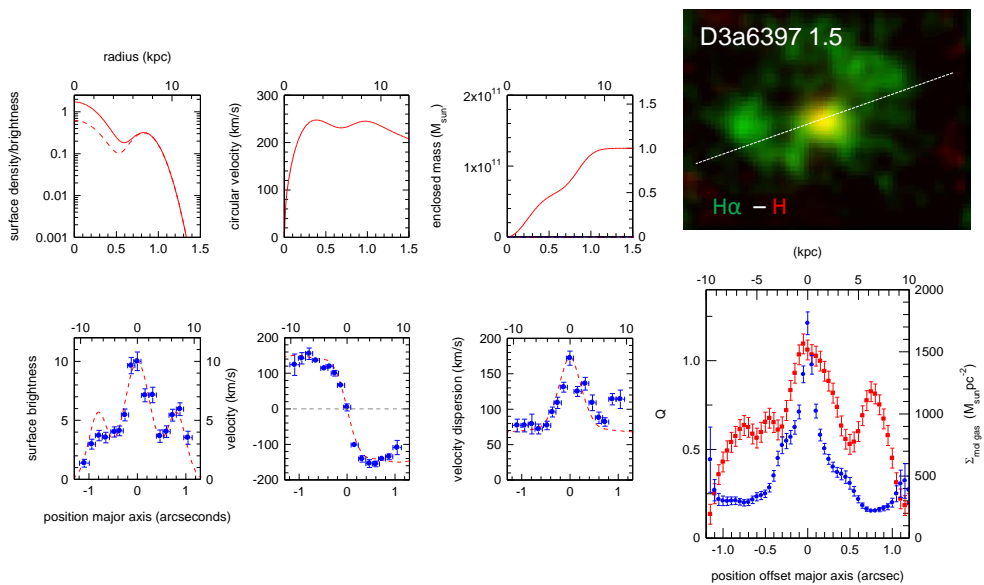

Figure A12.

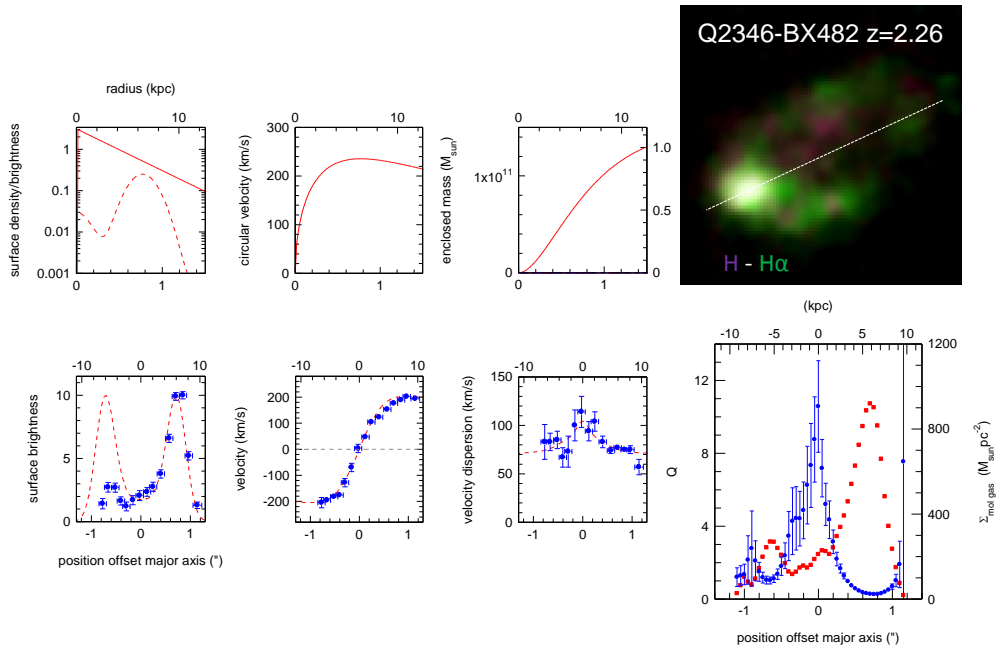

Figure A13.

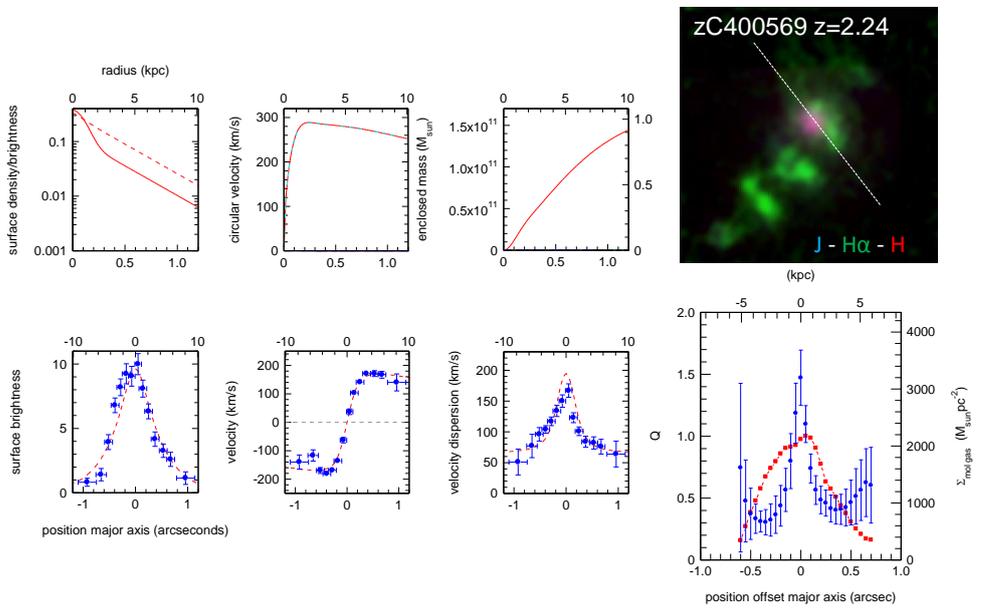

Figure A14.

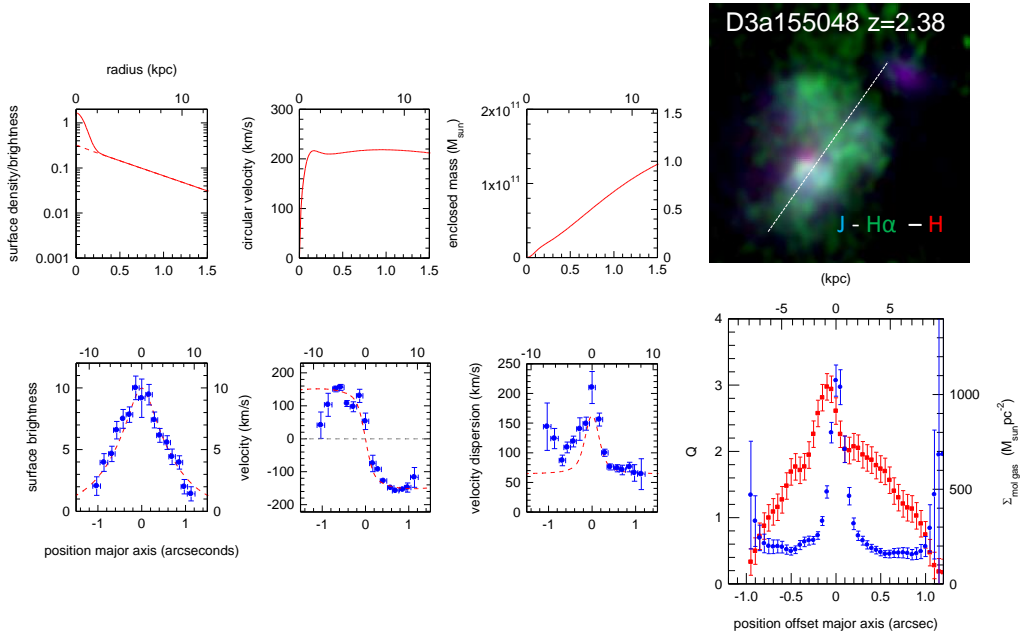

Figure A15.

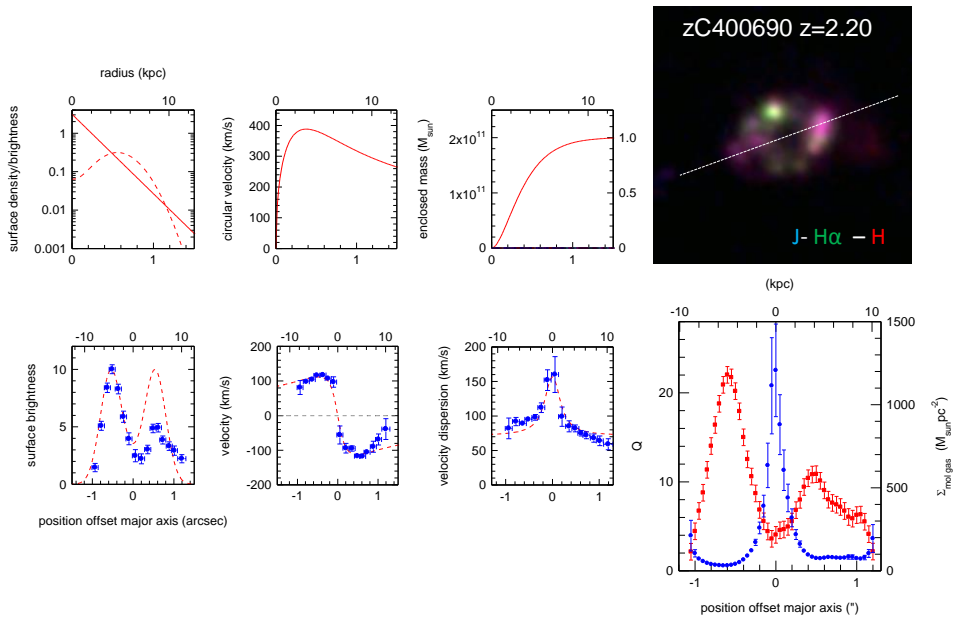

Figure A16.

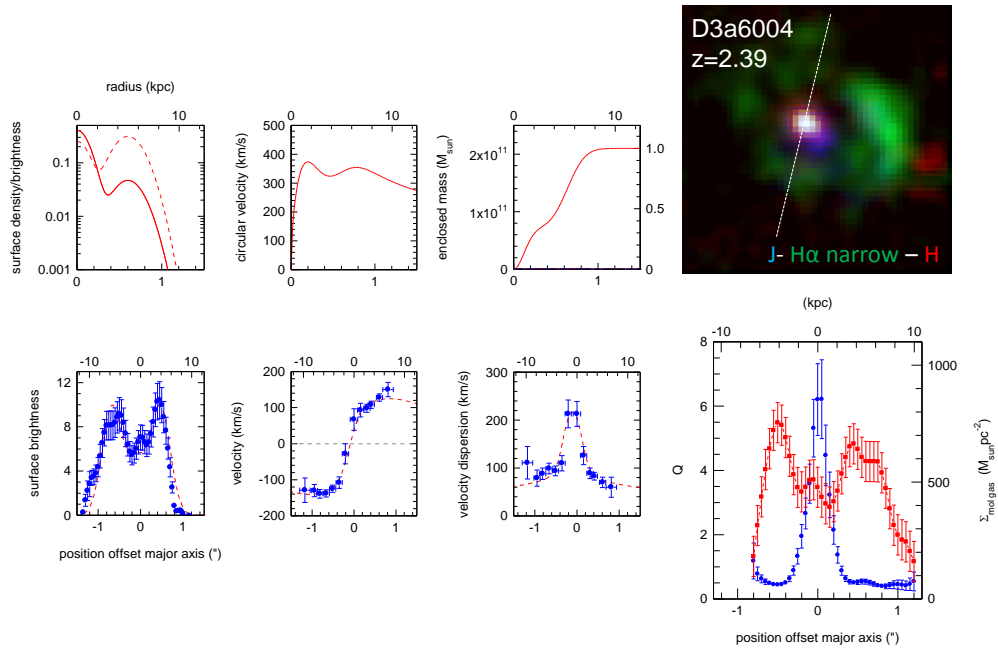

Figure A17.

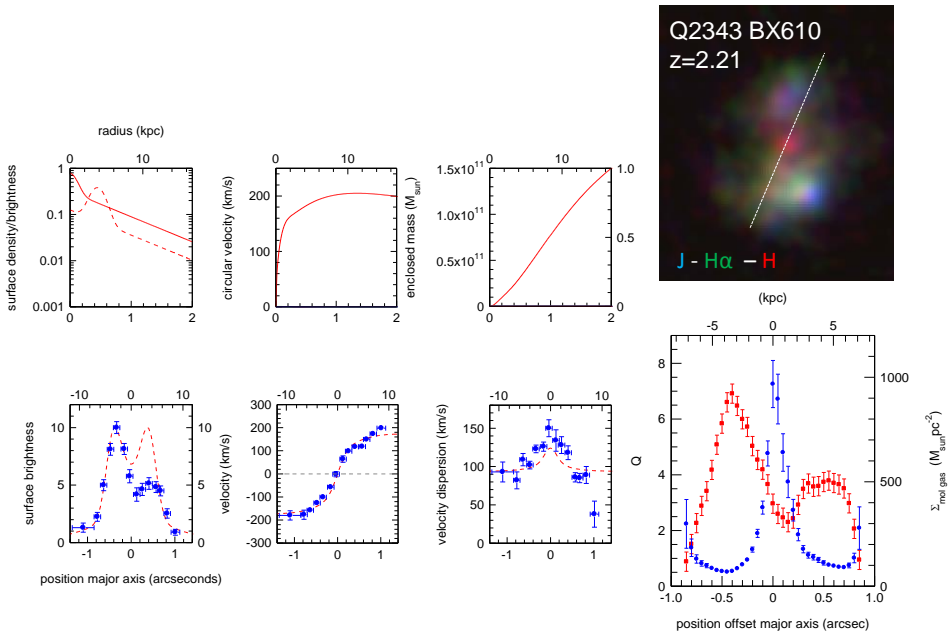

Figure A18.

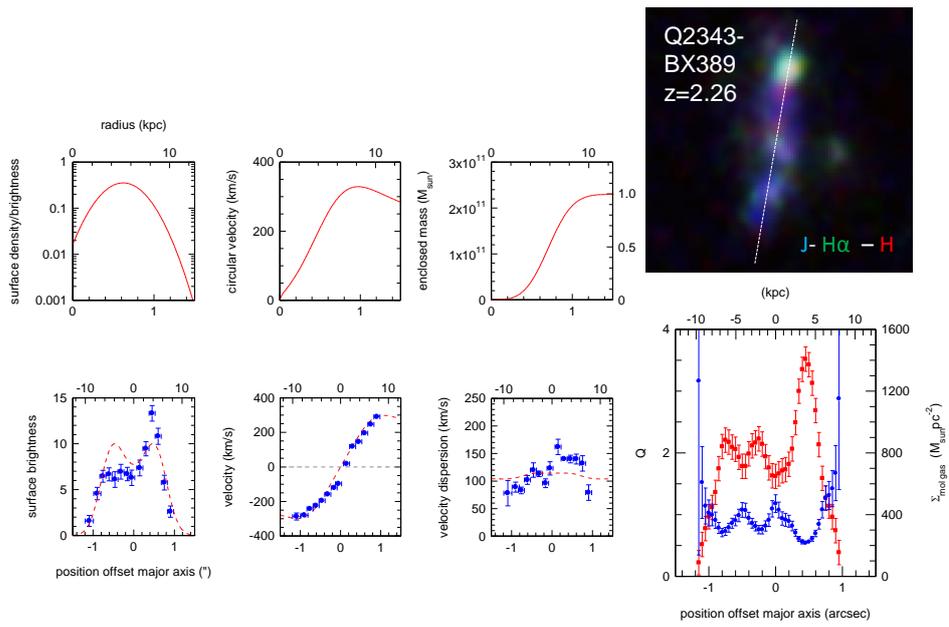

Figure A19.

# Table 1. properties of the galaxies

| galaxy | sSFR | R1/2 | del-R1/2 | logM* | log(M_dyn) | del-logM_dyn | mass-model | Hα distribution | κ_inner | log$\Sigma_{molgas}$inner | log$\Sigma_*$inner | Q_inner | del-Q | Q_outer | del-Q |
|---|---|---|---|---|---|---|---|---|---|---|---|---|---|---|---|
| | Gyr-1 | kpc | kpc | Msun | Msun | Msun | | | km/s/kpc | Msun/pc^2 | Msun/pc^2 | | | | |
| zC415876 | 7.0 | 1.79 | 0.31 | 9.96 | 10.36 | 0.15 | exponential | exponential | 391 | 3.20 | 2.40 | 0.98 | 0.10 | 0.49 | 0.07 |
| BX455 | 3.7 | 2.70 | 0.46 | 10.00 | 10.40 | 0.10 | exponential | exponential | 400 | 3.04 | | 1.40 | 0.09 | 0.70 | 0.13 |
| zC403741 | 1.5 | 2.50 | 0.43 | 10.64 | 10.50 | 0.17 | exponential | ring + exponential | 400 | 2.65 | | 1.78 | 0.12 | 0.85 | 0.06 |
| GK2363 | 3.2 | 2.50 | 0.75 | 9.97 | 10.52 | 0.14 | exponential | exponential | 400 | 3.21 | | 0.54 | 0.06 | 0.29 | 0.04 |
| zC405226 | 8.4 | 4.40 | 0.75 | 9.97 | 10.58 | 0.11 | exponential | exponential | 190 | 2.85 | 2.54 | 1.29 | 0.26 | 0.43 | 0.14 |
| zC405501 | 11.1 | 7.70 | 1.31 | 9.92 | 10.68 | 0.11 | ring + bulge | ring + bulge | 125 | 2.56 | 2.53 | 1.29 | 0.14 | 0.18 | 0.01 |
| zC410041 | 11.3 | 5.47 | 0.94 | 9.66 | 10.73 | 0.18 | exponential | exponential + ring | 220 | 2.72 | | 1.57 | 0.35 | 0.30 | 0.06 |
| zC410123 | 12.6 | 4.8 | 0.82 | 9.62 | 10.78 | 0.18 | exponential | exponential + ring | 230 | 2.73 | 2.71 | 1.85 | 0.28 | 0.60 | 0.08 |
| zC400528 | 1.4 | 1.57 | 0.27 | 11.00 | 10.85 | 0.21 | exponential | exponential + ring | 650 | 3.45 | 3.37 | 0.75 | 0.05 | 0.38 | 0.08 |
| GK2540 | 0.8 | 11.2 | 1.90 | 10.28 | 10.93 | 0.25 | exponential | ring + exponential | 230 | 1.27 | | 5.27 | 3.00 | 1.58 | 0.39 |
| zC407302 | 12.3 | 4.60 | 0.78 | 10.38 | 11.04 | 0.25 | exponential | exponential | 460 | 3.43 | 2.73 | 0.85 | 0.03 | 0.38 | 0.08 |
| D3a6397 | 3.8 | 6.20 | 1.05 | 11.08 | 11.15 | 0.21 | bulge + ring | ring + bulge | 550 | 3.17 | | 0.91 | 0.10 | 0.18 | 0.02 |
| BX482 | 4.9 | 5.48 | 0.94 | 10.30 | 11.20 | 0.24 | exponential | ring (+expon.) | 500 | 2.27 | 2.41 | 6.82 | 1.07 | 0.61 | 0.13 |
| zC400569 | 1.4 | 5.70 | 0.97 | 11.08 | 11.26 | 0.10 | exponential + bulge | exponential (or ring) | 900 | 3.39 | 4.05 | 1.25 | 0.21 | 0.37 | 0.08 |
| D3a15504 | 1.5 | 6.70 | 1.14 | 10.89 | 11.28 | 0.10 | exponential + bulge | exponential | 700 | 2.94 | 3.49 | 2.35 | 0.16 | 0.52 | 0.04 |
| zC406690 | 5.3 | 5.52 | 0.94 | 10.60 | 11.34 | 0.14 | exponential | ring (+expon.) | 1062 | 2.41 | 2.23 | 12.13 | 3.00 | 0.83 | 0.09 |
| D3a6004 | 1.4 | 5.60 | 0.95 | 11.48 | 11.36 | 0.16 | bulge +ring | ring + bulge | 1000 | 2.68 | 4.06 | 5.16 | 0.96 | 0.51 | 0.07 |
| BX610 | 0.8 | 4.90 | 0.83 | 11.00 | 11.38 | 0.14 | exponential + bulge | ring + exponential | 450 | 2.62 | 3.26 | 5.27 | 0.62 | 0.70 | 0.08 |
| BX389 | 2.7 | 6.80 | 1.16 | 10.60 | 11.41 | 0.18 | ring | ring | 103 | 2.84 | 2.45 | 0.99 | 0.09 | 0.81 | 0.07 |